\documentclass[a4paper,aps,prd,twocolumn,showpacs,superscriptaddress,longbibliography,nofootinbib]{revtex4-1}
\pdfoutput=1

\usepackage{enumerate}
\usepackage{graphicx}
\usepackage[caption=false]{subfig}
\usepackage{hyperref}

\usepackage{bm}
\usepackage{amsmath}
\usepackage{amssymb}

\newcommand{\USFT}{^{\textrm{SFT}}}
\newcommand{\Udet}{^{\textrm{det}}}
\newcommand{\Ueff}{^{\textrm{eff}}}
\newcommand{\Umax}{^{\textrm{max}}}
\newcommand{\Umid}{^{\textrm{mid}}}
\newcommand{\Upppr}{^{\,\prime\prime\prime}}
\newcommand{\Uppr}{^{\,\prime\prime}}
\newcommand{\Upr}{^{\,\prime}}
\newcommand{\Ustart}{^{\textrm{start}}}
\newcommand{\Ustop}{^{\textrm{stop}}}
\newcommand{\Zn}[1][n]{\mathbb{Z}^{#1}}
\newcommand{\bbR}{\mathbb{R}}
\newcommand{\calA}{\mathcal{A}}
\newcommand{\calB}{\mathcal{B}}
\newcommand{\calF}{\mathcal{F}}
\newcommand{\calN}{\mathcal{N}}
\newcommand{\calO}{\mathcal{O}}
\newcommand{\calV}{\mathcal{V}}
\newcommand{\cohsmat}[2][\ell]{\cohs[#1]{\mat{#2}}}
\newcommand{\cohsvec}[2][\ell]{\cohs[#1]{\vec{#2}}}
\newcommand{\cohs}[2][\ell]{\coh{#2}_{#1}}
\newcommand{\coh}[1]{\widetilde{#1}}
\newcommand{\frakR}{\mathfrak{R}}
\newcommand{\lalapps}[1]{\textsf{lalapps\_#1}}
\newcommand{\mat}[1]{\boldsymbol{\mathrm{#1}}}
\newcommand{\ndot}[2][s]{#2^{(#1)}}
\newcommand{\octave}[1]{\textsf{#1.m}}
\newcommand{\pluseq}{\mathrel{+}=}
\newcommand{\semimat}[1]{\semi{\mat{#1}}}
\newcommand{\semivec}[1]{\semi{\vec{#1}}}
\newcommand{\semi}[1]{\widehat{#1}}
\newcommand{\sigvec}[1]{\sig{\vec{#1}}}
\newcommand{\sig}[1]{#1^\mathrm{s}}
\newcommand{\uES}{_{\textrm{ES}}}
\newcommand{\uE}{_{\textrm{E}}}
\newcommand{\uFstat}{_{\calF}}
\newcommand{\uSGL}{_{\textrm{SGL}}}
\newcommand{\ua}{_{a}}
\newcommand{\ub}{_{b}}
\newcommand{\ucache}{_{\textrm{cache}}}

\newcommand{\uc}{_{c}}
\newcommand{\ufid}{_{\textrm{fiducial}}}
\newcommand{\uinj}{_{\textrm{inj}}}
\newcommand{\uiter}{_{\textrm{iter}}}
\newcommand{\umax}{_{\textrm{max}}}
\newcommand{\umean}{_{\textrm{mean}}}
\newcommand{\umin}{_{\textrm{min}}}
\newcommand{\uother}{_{\textrm{other}}}
\newcommand{\uoutput}{_{\textrm{out}}}
\newcommand{\upad}{_{\textrm{pad}}}
\newcommand{\uquery}{_{\textrm{query}}}
\newcommand{\useg}{_{\ell}}
\newcommand{\usemilogtenbsgl}{_{\semi\calB\uSGL}}
\newcommand{\usemimeantwof}{_{2\calF\umean}}
\newcommand{\usemisegsumtwof}{_{2\calF\usum}}
\newcommand{\usky}{_{\textrm{sky}}}
\newcommand{\usum}{_{\textrm{sum}}}
\newcommand{\utotal}{_{\textrm{tot}}}
\newcommand{\utwoF}{_{2\calF}}
\renewcommand{\vec}[1]{\boldsymbol{#1}}

\newcommand{\gitDate}{2018-07-02 16:36:12 +1000}
\newcommand{\gitID}{cd3bd73}
\newcommand{\gitStatus}{CLEAN}

\begin{document}

\title{Implementing a semicoherent search for continuous gravitational waves using optimally-constructed template banks}
\author{K. Wette}
\email{karl.wette@anu.edu.au}
\affiliation{ARC Centre of Excellence for Gravitational Wave Discovery (OzGrav) and Centre for Gravitational Physics, Research School of Physics and Engineering, The Australian National University, ACT 0200, Australia}
\affiliation{Max Planck Institute for Gravitational Physics (Albert Einstein Institute), D-30167 Hannover, Germany}
\author{S. Walsh}
\affiliation{Department of Physics, University of Wisconsin, Milwaukee, WI 53201, USA}
\affiliation{Max Planck Institute for Gravitational Physics (Albert Einstein Institute), D-30167 Hannover, Germany}
\author{R. Prix}
\affiliation{Max Planck Institute for Gravitational Physics (Albert Einstein Institute), D-30167 Hannover, Germany}
\author{M. A. Papa}
\affiliation{Max Planck Institute for Gravitational Physics (Albert Einstein Institute), D-30167 Hannover, Germany}
\affiliation{Department of Physics, University of Wisconsin, Milwaukee, WI 53201, USA}

\date{\gitDate, commit \gitID-\gitStatus}

\begin{abstract}
All-sky surveys for isolated continuous gravitational waves present a significant data-analysis challenge.
Semicoherent search methods are commonly used to efficiently perform the computationally-intensive task of searching for these weak signals in the noisy data of gravitational-wave detectors such as LIGO and Virgo.
We present a new implementation of a semicoherent search method, \emph{Weave}, that for the first time makes full use of a parameter-space metric to generate banks of search templates at the correct resolution, combined with optimal lattices to minimize the required number of templates and hence the computational cost of the search.
We describe the implementation of Weave and associated design choices, and characterize its behavior using semi-analytic models.
\end{abstract}

\pacs{04.80.Nn, 95.55.Ym, 95.75.Pq, 97.60.Jd}

\maketitle

\section{Introduction}\label{sec:introduction}

The detections of short-duration gravitational-wave events from the inspiral and merger of binary black holes~\cite{LIGOVirg2016:ObsGrvWvBnBHMr,LIGOVirg2016:ObsGrvW22BnBHCl:GW151226,LIGOVirg2017:Obs50SBnBHClRd02:GW170104,LIGOVirg2017:Obs19SlBnBlHCls:GW170608,LIGOVirg2017:ThrObGrvWBnBHCl:GW170814} and binary neutron stars~\cite{LIGOVirg2017:ObsGrvWBnNtSIn:GW170817} are enabling advances across astronomy, astrophysics, and cosmology.
As the gravitational-wave detectors LIGO~\cite{LIGO2009:LILsIntGrvOb,LIGO2015:AdvLIG}, Virgo~\cite{Virg2015:AdVScnIntGrWDt} improve in sensitivity in the coming years, and as new detectors KAGRA~\cite{KAGR2012:DtCnfKAJCrGrvD} and LIGO India~\cite{Unni2013:ILgISPGrvWRsPrMtI} come online, it may become possible to detect gravitational radiation from other astrophysical phenomena.
Rapidly-spinning, non-axisymmetrically-deformed neutron stars will emit gravitational waves in the form of continuous quasi-sinusoidal signals, and remain an intriguing prospect for detection with advanced instruments.
Searches for continuous gravitational waves in contemporary LIGO and Virgo data are ongoing~\cite[e.g.][]{LIGOVirg2017:FrSrGrvWKPlAdvL,LIGOVirg2017:SrGrvWScXFAdLObsRHMM,LIGOVirg2017:AlSrPrdGrvWvOLD,LIGOVirg2017:ULGrvWScXMdCrsSAdLD,LIGOVirg2017:FLwfEnASCnGrvWAdLD}.

Since the maximum non-axisymmetric deformation of neutron stars is expected to be small~\cite[e.g.][]{JohnOwen2013:MxElDfrRltSt}, continuous waves are expected to be weak relative to the sensitivity of the current generation of interferometric detectors.
Consequentially there has accumulated a significant body of research devoted to the data analysis challenge of extracting such weak signals from the gravitational-wave detector data.
Early results~\cite{BradEtAl1998:SrcPrdSrLI,JaraEtAl1998:DAnGrvSgSpNSSDtc} focused on the method of matched filtering the entire dataset against the known continuous-wave signal model; while theoretically optimal (in the Neyman–Pearson sense), this method quickly becomes computationally intractable if some or all of the model parameters are unknown.
Such is the case if one wished to target an interesting sky direction e.g.\ associated with a supernova remnant~\cite[e.g.][]{WettEtAl2008:SrGrvWvCssLI} or a low-mass X-ray binary~\cite[e.g.][]{LIGOVirg2017:SrGrvWScXFAdLObsRHMM,LIGOVirg2017:ULGrvWScXMdCrsSAdLD}, or perform an all-sky survey for isolated continuous-wave sources unassociated with known pulsars~\cite[e.g.][]{BradEtAl1998:SrcPrdSrLI}.
It is the latter type of search that is the subject of this paper.

The additional challenge of a practical upper limit on the computational cost of all-sky searches has spurred the development of various sub-optimal but computationally-tractable \emph{hierarchical} or \emph{semicoherent} algorithms~\cite{BradCrei2000:SrcPrSrLIIHrrSr:II}.
They share a common approach: the dataset (which for this example we assume is contiguous) with timespan $\semi T$ is partitioned into $N$ segments, each with timespan $\coh T$.
A fully-coherent matched filter search is then performed individually for each segment.
Most~\footnote{A few methods instead look for significant templates which are coincident between segments~\cite{LIGO2009:EnsSrPrGrvWvLSD,PoghEtAl2015:ArImpPrlSfSPGrWS}} methods then combine segments by incoherently summing the power from $N$ filters, one from each segment, which together follow a consistent frequency evolution as dictated by the continuous-wave signal model.
The phase evolution need not be continuous over the $N$ filters, however; nor need the gravitational-wave amplitudes in each segment be consistent.
This loss of complete signal self-consistency comes, however, with a computational benefit: while the computational cost of a fully-coherent matched filter search of the entire dataset scales as $\semi T^n = N^n \coh T^n$ with $n$ a high power~$\sim 5$ to~6, the cost of a semicoherent method typically scales as $N^m \coh T^n$ with $m \sim 2 \ll n$~\cite{PrixShal2012:SCntGrvWOpStMFCmC}.
The strain sensitivities of a fully-coherent and semicoherent search typically scale as $N^{1/2} \coh T^{1/2}$ and $N^{1/4w} \coh T^{1/2}$ respectively, with $w \ge 1$~\cite{PrixShal2012:SCntGrvWOpStMFCmC,Wett2012:EsSnWdpSrGrvP}; for the loss of a factor $N^{\sim 1/4}$ in sensitivity, a semicoherent method is able to gain by being able to analyze large (e.g.\ $\semi T \gtrsim 1$~year) datasets, whereas a fully-coherent search would be computationally restricted to a much shorter (e.g.\ $\semi T \ll 1$~year) subset.

An important early advance in the development of semicoherent methods was the adaption of the Hough transform~\cite{Houg1959:McAnlBbChPct}, originally created to analyze tracks in bubble chamber photographs, to instead track the frequency evolution of a continuous gravitational-wave signal~\cite{SchuPapa1999:EnAlHrrASrLngGSG600}.
A number of variations of the Hough transform have been implemented, which map the signal track in the time--frequency plane to either its sky position at a fixed reference frequency and frequency derivative~\cite{KrisEtAl2004:HgTrSrCntGrvW}, or conversely to its reference frequency and frequency derivative at a fixed sky position~\cite{AntoEtAl2008:DtcPrGrvWSrHTrFVFP,AstoEtAl2014:MASrCnGrvWSUFrqTr}.
The detection statistic computed, the \emph{number count}, sums either zero or one from each segment depending on whether the significance of a filter exceeds a set threshold.
Some variations use short-duration ($\coh T \sim 1800$s) segments and incoherently sum power above threshold from each segment; others analyze longer segments, and set a threshold on the \emph{$\cal F$-statistic}~\cite{JaraEtAl1998:DAnGrvSgSpNSSDtc} which computes the matched filter analytically maximized over the gravitational-wave amplitudes.
Another modification is to weigh each segment by the antenna response function of the detector, and to sum these weights instead of zero or one~\cite{PaloEtAl2005:AdpHTrnSrPrSr,KrisSint2007:HgSrImpSns}.

Two semicoherent methods which use short-duration segments but which, unlike the Hough transform methods, sum power without thresholding are the StackSlide~\cite{MendLand2005:StcHgSrSNStt} and PowerFlux~\cite{Derg2010:DscPw2AlImp} methods.
The StackSlide method builds a time--frequency plane, where each column represents a segment.
For each choice of signal parameters, it ``slides'' each column up and down in frequency so that a signal with those parameters would follow a horizontal line, and then ``stacks'' (i.e.\ sums) the columns horizontally to accumulate the signal power over time for each frequency bin.
(Due to this intuitive representation of a semicoherent search method, the term \emph{StackSlide} is often used to refer to semicoherent methods in general~\cite[e.g.][]{PrixShal2012:SCntGrvWOpStMFCmC}.)
The PowerFlux method follows a similar methodology, and in addition weights the power from each segment by that segment's noise level and antenna response function, so that segments containing transient instrumental noise and/or where the response of the detector is weak are deweighted.
A ``loosely coherent'' adaption to PowerFlux allows the degree of phase consistency imposed at the semicoherent stage to be controlled explicitly~\cite{Derg2010:BSrcNDmnSgLChAp,Derg2012:LsChrSrSWllSg}.
A third semicoherent method~\cite{PletAlle2009:ExLrCrrDCnGrvW,Plet2010:PrmMSmSrCnGrvW} was developed based on the observance of global correlations between search parameters~\cite{Plet2008:PrmCrOStCnGrvD} and uses longer segments analyzed with the $\calF$-statistic.
A comprehensive comparison of many of the all-sky search methods described above is performed in~\cite{WalsEtAl2016:CmpMtDtGrvWUnNS}.

Aside from developments in semicoherent search techniques, two other ideas have played an important role in the development of continuous gravitational-wave data analysis.
First is the use of a \emph{parameter-space metric}~\cite{BalaEtAl1996:GrvWClsBDStMCEsPr,Owen1996:STmGrvWInsBnCTmS,BradEtAl1998:SrcPrdSrLI}, which is used to determine the appropriate resolution of the bank of template signals such that the \emph{mismatch}, or fractional loss in signal-to-noise ratio between any signal present in the data and its nearest template, never exceeds a prescribed maximum.
The metric of the $\calF$-statistic for continuous-wave signals was first studied rigorously in~\cite{Prix2007:SrCnGrvWMMltFs}.
An approximate form of the metric was utilized in semicoherent search methods developed by~\cite{AstoEtAl2002:DAnlGrvSgSpNtSIVAS:IV}, and a related approximation was used in~\cite{PletAlle2009:ExLrCrrDCnGrvW,Plet2010:PrmMSmSrCnGrvW}.
The latter approximation, however, lead to an underestimation of the number of required templates in the sky parameter space when analyzing long data stretches; an improved approximate metric developed in~\cite{WettPrix2013:FPrmMtASrGrvPl,Wett2015:PrmMASmSrGrvPl} addresses this limitation.
It was also later realized that a further approximation fundamental to the metric derivation -- namely that the prescribed maximum mismatch (as measured by the metric) could be assumed small -- generally does not hold under realistic restrictions on computational cost.
This issue was addressed in~\cite{Wett2016:EmExRVlPrmMASGrvP} which computed an empirical relation between the metric-measured mismatch and the true mismatch of the $\calF$-statistic.

A second important idea is the borrowing of results from lattice theory~\cite[e.g.][]{ConwaySloane1988} to optimize the geometric placement of templates within the search parameter space, so as to fulfill the maximum prescribed mismatch criteria described above with the smallest possible density of templates~\cite{OwenSath1999:MFlGrvWInCBCmpCTPl,JaraKrol2005:GrvDAnFrSAppGsC}.
Practical algorithms for generating template banks for continuous-wave searches, using both the parameter-space metric and optimal lattices, were proposed in~\cite{Prix2007:TmpSrGrvWEfLCFPrS,Wett2014:LTmPlcChASrGrvP}.
An alternative idea studied in~\cite{MessEtAl2009:RnTmpBRlLtCvr,MancVall2010:CAIMtrMtrPlRStTmB} is to instead place templates at random, using the parameter-space metric only as a guide as to the relative density of templates; this idea has found utility in searches for radio~\cite[e.g.][]{KnisEtAl2013:EnsDsc24PlPrMlPS} and X-ray~\cite[e.g.][]{MessPatr2015:SmcSrWkPlsAQX} pulsars.

The number of computations that must be performed during an all-sky search, even when utilizing an efficient semicoherent search method, remains formidable.
For example, a recent all-sky search~\cite{LIGOVirg2017:FLwfEnASCnGrvWAdLD} of data from the first Advanced LIGO observing run divided the data into $N = 12$ segments of timespan $\coh T = 210$~hours, performed $\sim 3{\times}10^{15}$ matched-filtering operations per segment, and finally performed $\sim 3{\times}10^{17}$ incoherent summations to combine filter power from each segment.
The total computational cost of the search was $\sim 6{\times}10^{5}$~CPU days, although this was distributed over $\calO(10^{4})$ computers volunteered through the Einstein@Home distributed computing project~\cite{EinsteinAtHome}.
Nevertheless, the significant number of filtering/incoherent summation operations that must be performed during a typical all-sky search emphasizes the need to optimize the construction of the template banks, and thereby minimize the computational cost of the search, as much as practicable.

In this paper we present \emph{Weave}, an implementation of a semicoherent search method for continuous gravitational waves.
This implementation brings together, for the first time, several strands of previous research: the use of a semicoherent method to combine data segments analyzed with the $\calF$-statistic, combined with optimal template placement using the parameter-space metric of~\cite{WettPrix2013:FPrmMtASrGrvPl,Wett2015:PrmMASmSrGrvPl} and optimal lattices~\cite{Wett2014:LTmPlcChASrGrvP}.
After a review of relevant background information in Section~\ref{sec:background}, the Weave implementation is presented in Section~\ref{sec:weave-implementation}.
In Section~\ref{sec:models-weave-behav} we demonstrate that important behaviors of the Weave implementation can be modeled semi-analytically, thereby enabling characterization and optimization of a search setup without, in the first instance, the need to resort to time-consuming Monte-Carlo simulations.
In Section~\ref{sec:discussion} we discuss ideas for further improvement and extension.

\section{Background}\label{sec:background}

This section presents background material pertaining to the continuous-wave signal model, parameter-space metric, and template bank generation.

\subsection{Continuous-wave signals}\label{sec:cont-wave-sign}

The phase of a continuous-wave signal $\phi(t, \vec\lambda)$ at time $t$ at the detector is given by, neglecting relativistic corrections~\cite{JaraEtAl1998:DAnGrvSgSpNSSDtc},
\begin{equation}
\label{eq:phase-def}
\frac{ \phi(t, \vec\lambda) }{2\pi} \approx \sum_{s=0}^{s\umax} \ndot f \frac{ (t-t_0)^{s+1} }{ (s+1)! }
+ \frac{ \vec r(t) \cdot \vec n }{c} f\umax \,.
\end{equation}
The first term on the right-hand side primarily~\footnote{The rate of spindown observed at the Solar System barycenter is strictly a combination of the spindown observed in the source frame and the motion of the source~\cite{JaraEtAl1998:DAnGrvSgSpNSSDtc}; the latter is usually assumed to be small.} encodes the loss of rotational energy of the neutron star as observed from the Solar System barycenter: $f^{0}$ is the gravitational-wave frequency; and the \emph{spindowns} $f^{1}$, $f^{2}$, etc.\ are the 1st-order, 2nd-order, etc.\ rates of change of the gravitational-wave frequency with time.
All $\ndot f$ parameters are given with respect to a reference time $t_0$.
The second term on the right-hand side describes the Doppler modulation of the gravitational waves due to the motion of an Earth-based detector: $\vec r(t)$ is the detector position relative to the Solar System barycenter, thereby including both the sidereal and orbital motions of the Earth; and $\vec n$ is a unit vector pointing from the Solar System barycenter to the continuous-wave source.
The value of $f\umax$ is chosen conservatively to be the maximum of $f(t) \equiv d\phi(t, \vec\lambda) / dt$ over the timespan of the analyzed data.

Together the \emph{phase evolution parameters} $\vec\lambda = ( \vec n, \ndot f )$ parameterize the continuous-wave signal template; additional \emph{amplitude parameters} $\vec\calA$ are analytically maximized over when computing the $\calF$-statistic~\cite{JaraEtAl1998:DAnGrvSgSpNSSDtc}.
In noise the $\calF$-statistic is a central $\chi^2$ statistic with 4 degrees of freedom; when in the vicinity of a signal, the noncentrality parameter $\coh\rho^2$ of the noncentral $\chi^2$ distribution scales as $\coh\rho^2 \propto h_0^2 T / S_h[f(t)]$, where $h_0$ is the gravitational-wave amplitude, $T$ the amount of analyzed data, and $S_h[f(t)]$ is the noise power spectral density in the vicinity of the signal frequency $f(t)$.

\subsection{Parameter-space metric}\label{sec:param-space-metr}

The parameter-space metric $\mat g$ of the $\calF$-statistic is defined by a 2nd-order Taylor expansion of the noncentrality parameter:
\begin{multline}
\label{eq:rhosqr}
\rho^2(\vec\calA, \sigvec\lambda; \vec\lambda) = \rho^2(\vec\calA, \sigvec\lambda; \sigvec\lambda) \times \\ \big[ 1 - g_{ij}(\vec\calA, \sigvec\lambda) \Delta\lambda_i \Delta\lambda_j \big] + \calO(\Delta\vec\lambda^3) \,,
\end{multline}
with implicit summation over $i, j$, and where
\begin{equation}
\label{eq:metric}
g_{ij}(\vec\calA, \sigvec\lambda) \equiv \frac{-1}{2 \rho^2(\vec\calA, \sigvec\lambda; \vec\lambda)} \left. \frac{\partial^2 \rho^2(\vec\calA, \sigvec\lambda; \vec\lambda)}{\partial\lambda_i \partial\lambda_j} \right|_{\vec\lambda = \sigvec\lambda} \,.
\end{equation}
Here $\rho^2(\vec\calA, \sigvec\lambda; \sigvec\lambda)$ is the noncentrality parameter of the $\calF$-statistic when perfectly matched to a signal with parameters $\sigvec\lambda$, and $\rho^2(\vec\calA, \sigvec\lambda; \vec\lambda)$ is the noncentrality parameter when computed at some mismatched parameters $\vec\lambda = \sigvec\lambda + \Delta\vec\lambda$.
The mismatch is defined to be
\begin{align}
\mu(\vec\calA, \sigvec\lambda; \vec\lambda) &\equiv 1 - \frac{ \rho^2(\vec\calA, \sigvec\lambda; \vec\lambda) }{ \rho^2(\vec\calA, \sigvec\lambda; \sigvec\lambda) } \\
&= g_{ij}(\vec\calA, \sigvec\lambda) \Delta\lambda_i \Delta\lambda_j \,.
\end{align}
A very useful approximation to Eq.~\eqref{eq:metric} is the \emph{phase metric}~\cite{BradEtAl1998:SrcPrdSrLI,AstoEtAl2002:DAnlGrvSgSpNtSIVAS:IV,Prix2007:SrCnGrvWMMltFs}; it discards the amplitude modulation of the signal, and thereby the dependence on the known parameters $\calA$, retaining only dependence on the phase evolution parameters:
\begin{equation}
\label{eq:phase-metric}
g_{ij}(\sigvec\lambda) \equiv \left\langle \frac{\partial \phi}{\partial \lambda_i} \frac{\partial \phi}{\partial \lambda_j} \right\rangle - \left\langle \frac{\partial \phi}{\partial \lambda_i} \right\rangle \left\langle \frac{\partial \phi}{\partial \lambda_j} \right\rangle \,.
\end{equation}

\subsection{Optimal template placement}\label{sec:optim-templ-plac}

Template placement using optimal lattices is an example of a \emph{sphere covering}~\cite[e.g.][]{ConwaySloane1988}: a collection of lattice-centered $n$-dimensional spheres of equal radius.
The radius is chosen to be the smallest value that satisfies the property that each point in the $n$-dimensional parameter space is contained in at least one sphere.
A lattice where the ratio of the volume of the sphere to the volume of a lattice cell is minimized generates a minimal sphere covering, i.e.\ the minimal number of points required to cover a parameter space, which is exactly the property desired for template banks.
(For example, in two dimensions the minimal sphere covering is generated by the hexagonal lattice.)
We identify the covering spheres with the \emph{metric ellipsoids} $g_{ij}(\sigvec\lambda) \Delta\lambda_i \Delta\lambda_j \le \mu\umax$, where $\mu\umax$ is the prescribed maximum; it follows that the radii of the covering spheres is $\sqrt{\mu\umax}$.
A matrix transform $\mat T$ can then be constructed~\cite{Wett2014:LTmPlcChASrGrvP} which takes integers in $\vec\xi \in \Zn$ to template parameters $\vec\lambda$ to generate the template bank:
\begin{equation}
\label{eq:tmpl-bank-T}
\lambda_i = T_{ij} \xi_j = B_{ik}\big[ \mat g(\sigvec\lambda) \big] L_{kj} \xi_j \,,
\end{equation}
where $\mat B$ is a function of the metric $\mat g(\sigvec\lambda)$, and $L$ is particular to the lattice being used.
If $\mat T$ is a lower triangular matrix, an efficient algorithm~\cite{Wett2014:LTmPlcChASrGrvP} can be found for generating the template bank.

\subsection{Reduced supersky metric}\label{sec:reduc-supersky-metr}

In order for Eq.~\eqref{eq:tmpl-bank-T} to preserve the sphere covering property, however, it must be independent of the template parameters $\sigvec\lambda$.
Since $\mat B$ is a function of the metric, we require a metric which is also independent of $\lambda$: $\mat g(\sigvec\lambda) \rightarrow \mat g$.
The phase metric of Eq.~\eqref{eq:phase-metric} is independent of the frequency and spindown parameters $\ndot f$, but retains a dependence on sky position parameters, e.g.\ $\mat g(\sigvec\lambda) \rightarrow \mat g(\alpha, \delta)$ in terms of right ascension $\alpha$ and declination $\delta$.
The question of how to derive a useful metric which is independent of the sky position parameters, i.e.\ $\mat g(\alpha, \delta) \rightarrow \mat g$, has stimulated numerous approaches~\cite[e.g.][]{JaraKrol1999:DAnGrvSSpNSIIAEstPr:II,AstoEtAl2002:DAnlGrvSgSpNtSIVAS:IV,PletAlle2009:ExLrCrrDCnGrvW}.
In~\cite{WettPrix2013:FPrmMtASrGrvPl}, a useful $\mat g$ is derived through the following procedure:
\begin{enumerate}[(i)]

\item
  $\mat g(\alpha, \delta)$ is expressed in terms of the 3 components of $\vec n = (n_x, n_y, n_z)$, instead of 2 parameters such as $(\alpha, \delta)$.
  The 3 components of $\vec n$ are taken to be independent; geometrically this is equivalent to embedding $\mat g(\alpha, \delta)$ into a 3-dimensional \emph{supersky} parameter space, instead of being restricted to the 2-sphere defined by $(\alpha, \delta)$.
  In the supersky parameter space, $\mat g$ is independent of the sky position parameters, i.e.\ we have the desired $\mat g(\alpha, \delta) \rightarrow \mat g$, but with the addition of a 3rd unwanted parameter-space dimension.

\item
  A linear coordinate transform $(n_x, n_y, n_z, \ndot f) \rightarrow (n\ua, n\ub, n\uc, \ndot \nu)$ is derived which satisfies: $\mat g$ is diagonal in the sky position parameters $(n\ua, n\ub, n\uc)$, i.e.\ $g_{n\ua n\ub} = g_{n\ua n\uc} = g_{n\ub n\uc} = 0$; $g_{n\ua n\ua} \gg g_{n\uc n\uc}$; and $g_{n\ub n\ub} \gg g_{n\uc n\uc}$.
  The last two properties imply that the metric ellipsoids are much longer along the $n\uc$ axis than along the $n\ua$ and $n\ub$ axes.
  In computing the coordinate transform, use is made of the well-known correlation between the sky and frequency/spindown parameters of the continuous-wave signal~\cite[e.g.][]{PrixItoh2005:GPrmCrChSCnGrvW,Plet2008:PrmCrOStCnGrvD}.
  The correlations arise because, on sufficiently short timescales, the change in phase due to the cyclical sidereal and orbital motions of the Earth may be Taylor expanded as linear, quadratic, etc.\ changes in phase with time, and thereby are equivalent to changes in the frequency ($\ndot[0] f \equiv f$), 1st spindown ($\ndot[1] f \equiv \dot f$), etc.\ parameters.

\item
  Since, in the new coordinates $(n\ua, n\ub, n\uc, \nu^{(s)})$ the mismatch $\mu$ is only weakly dependent on $g_{n\uc n\uc}$, a useful approximate metric $\mat g$ is found by discarding the $n\uc$ dimension.
  Geometrically this corresponds to projecting the 3-dimensional supersky parameter space and metric onto the 2-dimensional $n\ua$--$n\ub$ plane.
  The resultant \emph{reduced supersky} parameter-space metric $\mat g$ and associated coordinates $(n\ua, n\ub, \nu^{(s)})$ has reduced the sky parameter space dimensionality back to 2, while retaining the property that $\mat g$ is parameter-independent.

\end{enumerate}

\section{Weave Implementation}\label{sec:weave-implementation}

\begin{figure}
\centering
\includegraphics[width=\linewidth]{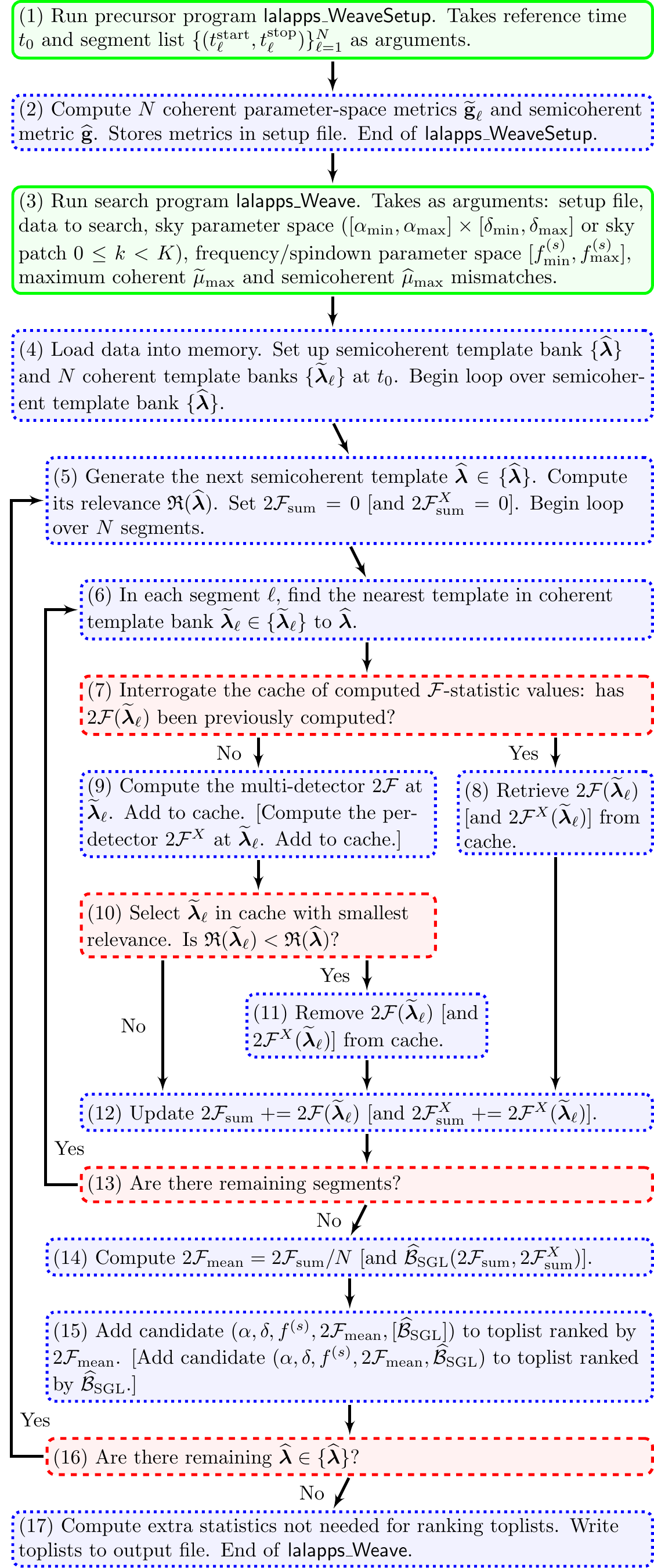}
\caption{\label{fig:weave_schematic}
Schematic of the Weave implementation.
Boxes with solid borders (green) represent actions taken by the user.
Boxes with dotted borders (blue) represent actions taken by the program.
Boxes with dashes borders (red) represent decisions the program must take.
Bracketed text denotes the optional computation of $\semi\calB\uSGL$.
See the text in Section~\ref{sec:weave-implementation} for a full description of the Weave implementation.
}
\end{figure}

\begin{figure}
\centering
\includegraphics[width=\linewidth]{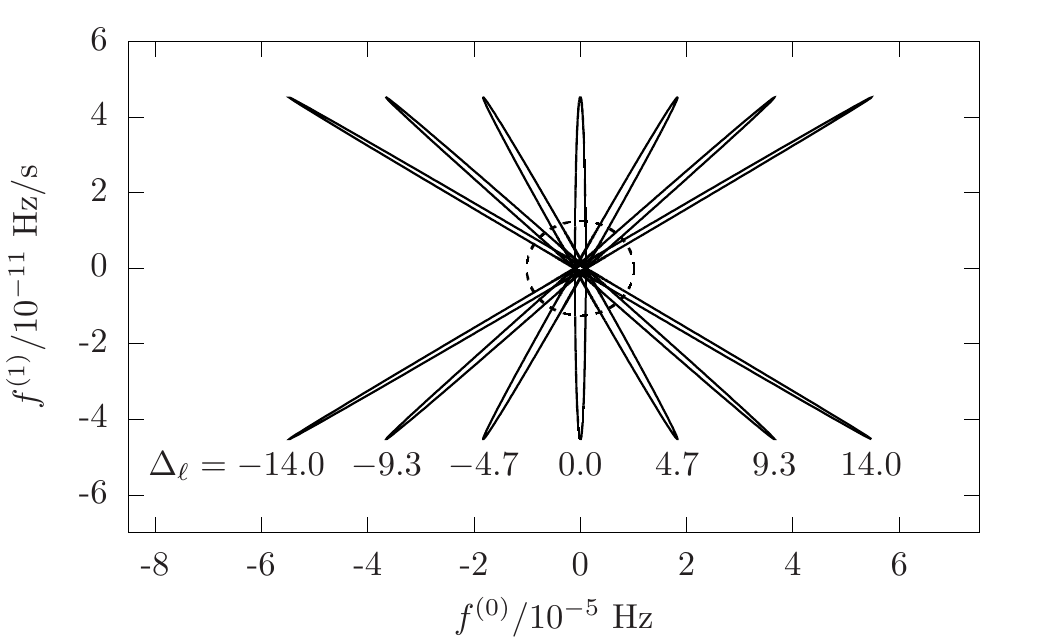}
\caption{\label{fig:freq_spin_templates}
Coherent (solid) and semicoherent (dashed) metric ellipses in the $(f, \dot f)$ plane.
The seven coherent metric ellipses are for segments with $\coh T = 2$~days, evenly spaced within a timespan of $\semi T = 30$~days; each ellipse is labeled by its value of $\Delta\useg$ / days; see Eq.~\eqref{eq:freq-spin-templates-delta}.
The reference time $t_0$ for all metrics is centered within $\semi T$.
The coherent and semicoherent metric ellipses are plotted at maximum mismatches of $\coh\mu\umax = 0.1$ and $\semi\mu\umax = 10$ respectively.
}
\end{figure}

\begin{figure*}
\centering
\includegraphics[width=\linewidth]{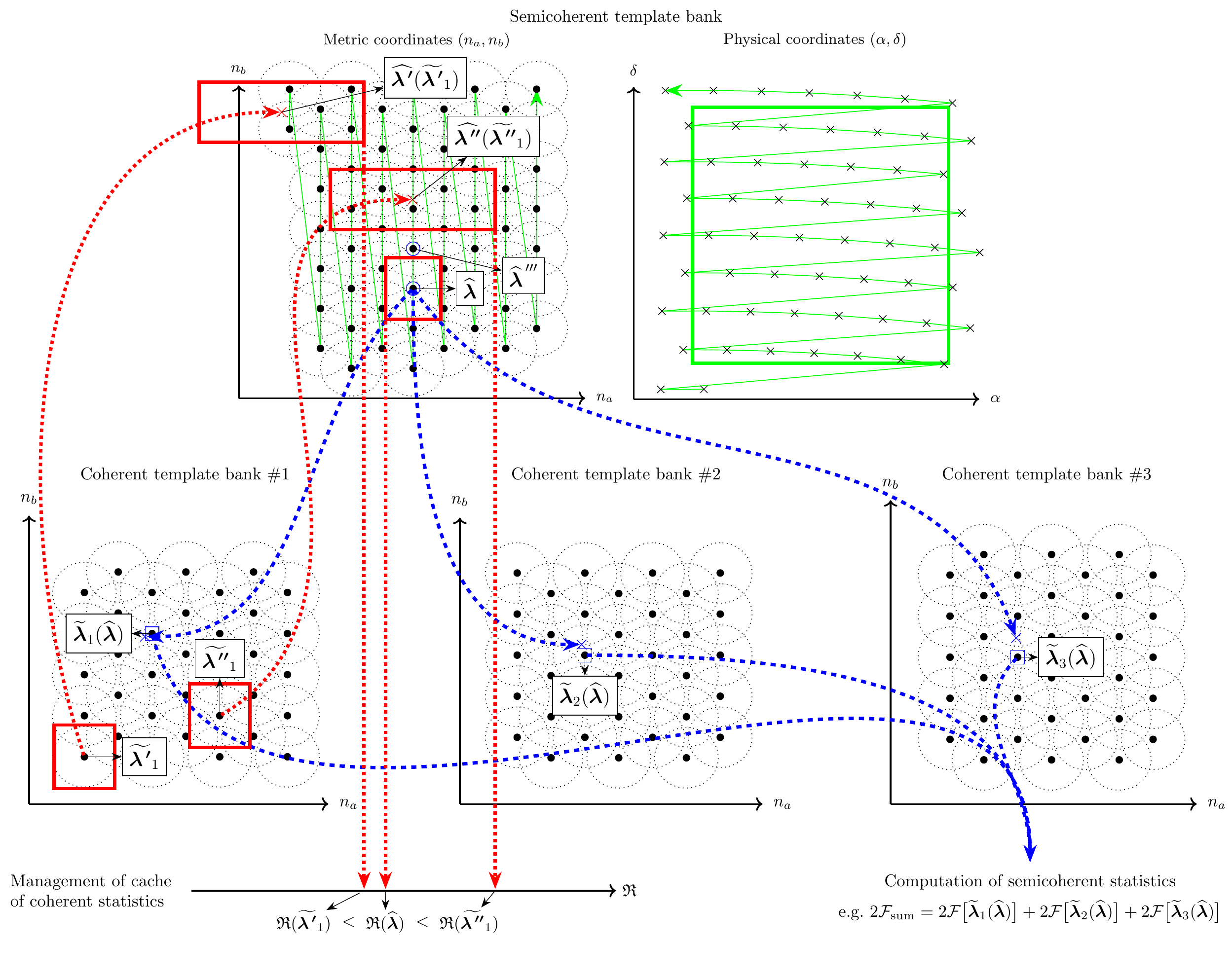}
\caption{\label{fig:weave_template_banks}
Diagram of the Weave template banks for an example search setup with 3 segments.
Shown are the semicoherent and coherent template banks in the two sky parameters of the reduced supersky metric $(n\ua, n\ub)$; the frequency and spindown dimensions are omitted.
For clarity the semicoherent and coherent template banks are plotted with metric ellipses of the same size.
Arrows with solid lines (green) represent iteration over the semicoherent template bank.
Arrows with dashed lines (blue) represent the process of computing semicoherent statistics, described in Section~\ref{sec:comp-semic-stat}.
Arrows with dotted lines (red) represent the process of managing the caches of coherent statistics, described in Section~\ref{sec:manag-cache-coher}.
Also shown is the semicoherent template bank in physical coordinates $(\alpha, \delta)$; the solid square (green) shows the rectangular boundary of the sky parameter space in these coordinates.
}
\end{figure*}

This section describes the Weave implementation of the semicoherent search method, a schematic of which is shown in Figure~\ref{fig:weave_schematic}.
The implementation is freely available as part of the LALSuite~\cite{LALSuite} gravitational-wave data analysis library.

\subsection{Overview}\label{sec:overview}

In step~1 the user runs a precursor program \lalapps{WeaveSetup}, which takes as an argument a list of $N$ segments $\{(t\Ustart\useg, t\Ustop\useg)\}_{\ell=0}^{N-1}$ into which the dataset is to be partitioned.
The program computes in step~2 the $N$ coherent parameter-space metrics $\cohsmat g$ used to construct template banks within each segment, and the semicoherent parameter-space metric $\semimat g$ used to incoherently combine segments.
The metrics are written to a \emph{setup file} in the FITS format~\cite{WellEtAl1981:FIFlxImTrnSy}.
Due to the numerical ill-conditionedness of the parameter-space metric~\cite{Prix2007:SrCnGrvWMMltFs,WettPrix2013:FPrmMtASrGrvPl}, this computation involves a bootstrapping process, whereby successively better-conditioned iterations of the supersky metric are computed, before then computing the reduced supersky metric as outlined in Section~\ref{sec:reduc-supersky-metr}.
Since this bootstrapping process can be time-consuming for large $N$, and may give slightly different results on different computer hardware, precomputing the metrics both saves computing time and adds robustness against numerical errors.
Note that, by Eq.~\eqref{eq:phase-def}, the sky components of the metrics will scale with $f\umax^2$; since its value depends on the search frequency parameter space, which is not known by \lalapps{WeaveSetup}, an arbitrary fiducial value $f\ufid$ is used, and the sky components of the metrics are later rescaled by $(f\umax / f\ufid)^2$.

In step~3 the user runs the main search program \lalapps{Weave}.
The principle arguments to this program are the setup file output by \lalapps{WeaveSetup}, the search parameter space, and the prescribed maximum mismatches $\coh\mu\umax$ and $\semi\mu\umax$ for the coherent and semicoherent template banks respectively.
The frequency and spindown parameter space is specified by ranges $[\ndot f\umin, \ndot f\umax]$, where $s = 0$, 1, etc.\ as required.
The sky search parameter space may be specified either as a rectangular patch in right ascension and declination $[\alpha\umin, \alpha\umax] \otimes [\delta\umin, \delta\umax]$, or alternatively partitioned into $K$ patches containing approximately equal number of templates (see Appendix~\ref{sec:prop-equal-area}), and a patch selected by an index $k$, $0 \le k < K$.
In step~4 various preparatory tasks are performed, such as loading the gravitational-wave detector data into memory, before beginning the main search loop.

The main search loop of a semicoherent search method may be structured in two complementary ways, which differ in the memory each requires to store intermediate results:
\begin{enumerate}[(i)]

\item
  The semicoherent template bank $\{\semivec\lambda\}$ is stored in memory, and the $N$ segments are processed in sequence.
  For each segment $\ell$, every coherent template $\cohsvec\lambda \in \{\cohsvec\lambda\}$ is mapped back to the semicoherent template bank, i.e.\ $\cohsvec\lambda \rightarrow \semivec\lambda(\cohsvec\lambda)$.
  Because the semicoherent template bank must track the continuous-wave signal over a larger timespan $\semi T \gg \coh T$ than the coherent template banks, it will contain a greater density of templates; the ratio of semicoherent to coherent template bank densities is the \emph{refinement factor} $\gamma \ge 1$~\cite{Wett2015:PrmMASmSrGrvPl,Plet2010:PrmMSmSrCnGrvW}.
  It follows that the mapping $\cohsvec\lambda \rightarrow \semivec\lambda(\cohsvec\lambda)$ will be one-to-many.

  As the $N$ segments are processed, any semicoherent detection statistic associated with $\semivec\lambda(\cohsvec\lambda)$ is then updated based on the corresponding coherent detection statistic associated with $\cohsvec\lambda$.
  For example, it is common to compute the summed $\calF$-statistic $2\calF\usum \equiv \sum_{\ell=0}^{N-1} 2\calF(\cohsvec\lambda)$; here we would then have $2\calF\usum\big[ \semivec\lambda(\cohsvec\lambda) \big] \pluseq 2\calF(\cohsvec[\ell]\lambda)$.
  Once every segment has been processed, computed $2\calF\usum$ for every $\semivec\lambda \in \{\semivec\lambda\}$ will exist in memory.
  The memory usage of the main search loop will therefore be proportional to the number of semicoherent templates $\semi\calN \equiv \gamma \times \langle\coh\calN\rangle$, where $\langle\coh\calN\rangle$ is the average number of templates in a coherent template bank.

\item
  The $N$ coherent template banks $\{\cohsvec\lambda\}$ are stored in memory, and the semicoherent template bank is processed in sequence.
  Each semicoherent template $\semivec\lambda \in \{\semivec\lambda\}$ is mapped back to the coherent template bank in each segment $\ell$, i.e.\ $\semivec\lambda \rightarrow \cohsvec\lambda(\semivec\lambda)$; since $\semi\calN \ge \coh\calN$ in each segment this mapping will be many-to-one.
  With these $N$ mappings in hand, the semicoherent detection statistics may be immediately computed in full, e.g.\ $2\calF\usum(\semivec\lambda) = \sum_{\ell=0}^{N-1} 2\calF\big[ \cohsvec\lambda(\semivec\lambda) \big]$.
  The memory usage of the main search loop will therefore be proportional to $N \times \langle\coh\calN\rangle$.

\end{enumerate}
For the parameter-space metric for all-sky searches, $\gamma \gg N$~\cite{Wett2015:PrmMASmSrGrvPl,Plet2010:PrmMSmSrCnGrvW}, and therefore the latter structuring given above will have the lower memory requirement; the Weave implementation uses this structuring of the main search loop.
The semicoherent template bank $\{\semivec\lambda\}$ is generated one template at a time using the algorithm described in~\cite{Wett2014:LTmPlcChASrGrvP}.
For each coherent template bank, an efficient lookup table~\cite{Wett2014:LTmPlcChASrGrvP} is constructed for the mapping $\semivec\lambda \rightarrow \cohsvec\lambda(\semivec\lambda)$.

We note an important distinction between the definition of the Weave template banks and the traditional StackSlide picture of a semicoherent search method.
In the latter picture, the frequency and spindown template banks of each segment are defined with respect to \emph{individual} reference times $(t_0)\useg$, typically the midtime of each segment.
When combining segments, therefore, the frequency and spindown parameters of each coherent template must be adjusted so as to bring the parameters of all segments to a common reference time $t_0$; this is the ``sliding'' step.
The Weave implementation, however, defines the frequency and spindown templates banks of all segments at the \emph{same} reference time $t_0$, which is also the reference time of the semicoherent bank.
Consequentially, there is no analogy to the ``sliding'' step of StackSlide.
Instead, the orientation of the metric ellipses in the $(f, \dot f)$ plane changes from segment to segment, as illustrated in Figure~\ref{fig:freq_spin_templates}.
As the absolute difference
\begin{equation}
\label{eq:freq-spin-templates-delta}
|\Delta\useg| = |t\Umid\useg - t_0|
\end{equation}
between the midtime of each segment $t\Umid\useg \equiv (t\Ustart\useg + t\Ustop\useg)/2$ and $t_0$ increases, both the extent of the ellipses in $f$ and the correlation between $f$ and $\dot f$ also increase.

Steps~5--16 comprise the main search loop; which performs two key tasks: the computation and output of the detection statistics over the semicoherent template bank (steps~5, 6, and 12--17), and the management of an internal cache of required detection statistics computed on each coherent template bank (steps~7--11).
These two tasks are described more fully in the following two sections, and with reference to a diagram of their operation in Figure~\ref{fig:weave_template_banks}.

In this section and in Figure~\ref{fig:weave_schematic} we focus for simplicity on the computation of the semicoherent $\calF$-statistics $2\calF\usum$ and $2\calF\umean \equiv 2\calF\usum / N$.
The computation of other detection statistics is also possible: in particular a family of Bayesian statistics has been developed which weigh the likelihood of a continuous wave signal against that of an instrumental line which appears in all segments~\cite{KeitEtAl2014:SCnGrvWImRbVInsA,KeitPrix2015:LnStCnGrvWSCUDSns}, or a transient instrumental line which appears only in one segment~\cite{Keit2016:RSmcSCnGrvWNSMInHDLTr}.
Computation of the former statistic, denoted $\semi\calB\uSGL$, is also illustrated in Figure~\ref{fig:weave_schematic}; it takes as input the multi-detector $2\calF\usum$ which uses data from all gravitational-wave detectors, as well as the per-detector $2\calF^X\usum$ which are computed from each detector $X$ individually.

\subsection{Computation of semicoherent statistics}\label{sec:comp-semic-stat}

In steps~5 and~16 (Figure~\ref{fig:weave_schematic}), the main loop of the search method generates successive points $\semivec\lambda$ in the semicoherent template bank.
An example of such a point is indicated in Figure~\ref{fig:weave_template_banks}.
Next, in steps~6 and~13, each segment $\ell$ is visited and the mapping $\semivec\lambda \rightarrow \cohsvec\lambda(\semivec\lambda)$ is performed.
The mapping used by Weave is nearest-neighbor interpolation: the $\semivec\lambda$ is expressed in the coherent metric coordinates of the $\ell$th segment, and the nearest (with respect to the metric) coherent template in the respective bank $\cohsvec\lambda(\semivec\lambda)$ is determined.
If the template bank is constructed on a lattice, efficient algorithms exist for determining the nearest point~\cite[e.g.][and references therein]{Wett2014:LTmPlcChASrGrvP}.
In Figure~\ref{fig:weave_template_banks}, example nearest coherent templates are labeled $\cohsvec[1]\lambda(\semivec\lambda)$, $\cohsvec[2]\lambda(\semivec\lambda)$, and $\cohsvec[3]\lambda(\semivec\lambda)$.

As each nearest point is determined, the coherent $\calF$-statistic in the respective segment is computed (steps~7--11, see Section~\ref{sec:manag-cache-coher}), and the value of the semicoherent statistic $2\calF\usum$ is updated (step~12).
Once all segments have been processed (step~13), additional semicoherent statistics such as $2\calF\umean$ are computed (step~14), and a \emph{candidate} comprising the signal parameters together with the computed semicoherent statistics is added (step~15) to one or more \emph{toplists} which ranks~\footnote{Toplists are implemented efficiently as a binary heap~\cite[e.g.][]{Morin2013}.} each candidate by a chosen semicoherent statistic.
The size of the toplists is generally of a fixed user-determined size so that only a fixed number of the most promising candidates will be returned.

Once the semicoherent template bank is exhausted (step~16) the toplists are written to an output file in the FITS format, and the search concludes (step~17).

\subsection{Management of cache of coherent statistics}\label{sec:manag-cache-coher}

It is important that the main search loop minimizes its memory usage as much as possible.
Even though in Section~\ref{sec:overview} we chose a structuring of the main search loop so as to reduce memory usage, a naive implementation which stores $N \times \langle\coh\calN\rangle$ coherent statistics would still require a prohibitive amount of memory, given that both $N$ and $\langle\coh\calN\rangle$ are typically large.
We therefore implement a per-segment cache which stores only those coherent statistics associated with coherent templates $\cohsvec\lambda$ accessible from the unprocessed portion of the semicoherent template bank via the mapping $\semivec\lambda \rightarrow \cohsvec\lambda(\semivec\lambda)$.
Put another way, if a $\cohsvec\lambda$ can no longer be mapped to by any $\semivec\lambda$ remaining in $\{\semivec\lambda\}$, then $2\calF(\cohsvec\lambda)$ can be safely removed from the cache.

In order to devise a cache management algorithm with the above desired properties, we first define an operator called \emph{relevance}, denoted $\frakR : \vec\lambda \rightarrow \bbR$.
The relevance operates on both coherent and semicoherent templates, and should satisfy the following property:
\begin{equation}
\label{eq:relevance-prop}
\parbox{0.85\linewidth}{For all $\cohsvec\lambda \in \{\cohsvec\lambda\}$ and for all $\semivec\lambda \in \{\semivec\lambda\}$, the condition $\frakR(\cohsvec\lambda) < \frakR(\semivec\lambda)$ implies that no mapping $\semivec\lambda \rightarrow \cohsvec\lambda(\semivec\lambda)$ exists in the remaining $\{\semivec\lambda\}$, and thus $2\calF(\cohsvec\lambda)$ can be safely removed from the cache.}
\end{equation}
A definition of $\frakR$ satisfying this property is derived as follows.

First, take any coherent template (e.g.\ $\cohsvec[1]{\lambda\Upr}$ in Figure~\ref{fig:weave_template_banks}) and surround it by its metric ellipsoid at mismatch $\coh\mu\umax$.
Then surround the metric ellipsoid in turn by its \emph{bounding box}, the smallest coordinate box which contains the ellipsoid~\cite[e.g.][]{Wett2014:LTmPlcChASrGrvP}; the metric ellipse bounding box centered on $\cohsvec[1]{\lambda\Upr}$ is also shown in Figure~\ref{fig:weave_template_banks}.
Now, transform the bounding box into the semicoherent parameter space; practically this may be achieved by expressing the coordinates of each vertex of the bounding box in the semicoherent metric coordinates.
See Figure~\ref{fig:weave_template_banks} for the transformed bounding box of $\cohsvec[1]{\lambda\Upr}$ in the semicoherent parameter space, which is centered on $\semivec{\lambda\Upr}(\cohsvec[1]{\lambda\Upr})$.

Note that, by definition, any semicoherent template $\semivec\lambda$ outside of the transformed bounding box centered on $\semivec{\lambda\Upr}(\cohsvec[1]{\lambda\Upr})$ \emph{cannot} map to $\cohsvec[1]{\lambda\Upr}$ under $\semivec\lambda \rightarrow \cohsvec\lambda(\semivec\lambda)$.
Thus, to determine whether $\cohsvec[1]{\lambda\Upr}$ is accessible by $\semivec\lambda$, we can compute whether $\semivec\lambda$ is within the transformed bounding box of $\semivec{\lambda\Upr}(\cohsvec[1]{\lambda\Upr})$.
To be conservative, however, we also surround $\semivec\lambda$ by its bounding box as shown in Figure~\ref{fig:weave_template_banks}, and instead compute whether the bounding boxes of $\semivec{\lambda\Upr}(\cohsvec[1]{\lambda\Upr})$ and $\semivec\lambda$ intersect.

To simplify the bounding box intersection calculation, we compare just the coordinates of the bounding boxes of $\semivec{\lambda\Upr}(\cohsvec[1]{\lambda\Upr})$ and $\semivec\lambda$ in \emph{one} dimension; for reasons that will soon be apparent, we choose the lowest-dimensional coordinate, $n\ua$.
First, we define the relevance $\frakR$ for both coherent and semicoherent templates:
\begin{subequations}
\label{eqs:relevance}
\begin{align}
\frakR(\cohsvec\lambda) &\equiv \parbox{0.65\linewidth}{the \emph{maximum} value of $n\ua$ within the transformed bounding box of $\cohsvec\lambda$,}
\intertext{and}
\frakR(\semivec\lambda) &\equiv \parbox{0.65\linewidth}{the \emph{minimum} value of $n\ua$ within the bounding box of $\semivec\lambda$.}
\end{align}
\end{subequations}
We now compute $\frakR(\cohsvec[1]{\lambda\Upr})$ and $\frakR(\semivec\lambda)$; in Figure~\ref{fig:weave_template_banks}, $\frakR(\cohsvec[1]{\lambda\Upr})$ is the $n_a$ coordinate of the right-most edge of the transformed bounding box of $\semivec{\lambda\Upr}(\cohsvec[1]{\lambda\Upr})$, and $\frakR(\semivec\lambda)$ is the $n_a$ coordinate of the left-most edge of the bounding box of $\semivec\lambda$.
In this example, $\frakR(\cohsvec[1]{\lambda\Upr}) < \frakR(\semivec\lambda)$, and it follows from the definition of $\frakR$ in Eqs.~\eqref{eqs:relevance} that the bounding boxes of $\semivec{\lambda\Upr}(\cohsvec[1]{\lambda\Upr})$ and $\semivec\lambda$ \emph{cannot} intersect.

On the other hard, let us choose another coherent template $\cohsvec[1]{\lambda\Uppr}$, and examine its relevance $\frakR(\cohsvec[1]{\lambda\Upr})$; here we have $\frakR(\cohsvec[1]{\lambda\Uppr}) > \frakR(\semivec\lambda)$ (see Figure~\ref{fig:weave_template_banks}).
From the simplified bounding box intersection calculation, we conclude that the bounding boxes of $\semivec{\lambda\Uppr}(\cohsvec[1]{\lambda\Uppr})$ and $\semivec\lambda$ \emph{could potentially} intersect, since at least in the $n_a$ dimension the bounding boxes overlap (although in this example the bounding boxes do not overlap in the $n_b$ dimension).

Finally, if for some $\semivec\lambda$ we have $\frakR(\cohsvec[1]{\lambda\Upr}) < \frakR(\semivec\lambda)$, then this condition is guaranteed to remain true for all remaining $\semivec\lambda$ in the template bank.
This is simply a consequence of the algorithm used to generate the semicoherent template bank~\cite{Wett2014:LTmPlcChASrGrvP}, which operates as follows: first, values of $n\ua$ are generated in a constant range $[{n\ua}\umin, {n\ua}\umax]$; then, for each value of $n\ua$, values of $n\ub$ are generated in ranges $[{n\ub(n\ua)}\umin, {n\ub(n\ua)}\umax]$ dependent on $n_a$, and so on.
It follows that the value of $n\ua$ can only increase during the generation of the semicoherent template bank, and since $\frakR(\semivec\lambda)$ is defined in terms of $n\ua$, it too can only increase.

To summarize, the relevance operator $\frakR$ defined by Eqs.~\eqref{eqs:relevance} satisfies the desired property given by Eq.~\eqref{eq:relevance-prop}.
In Figure~\ref{fig:weave_template_banks}, since $\frakR(\cohsvec[1]{\lambda\Upr}) < \frakR(\semivec\lambda)$, the cache management algorithm would discard any coherent statistics associated with $\cohsvec[1]{\lambda\Upr}$ from memory, since they cannot be accessed by $\semivec\lambda$ nor any remaining semicoherent template.
On the other hard, the algorithm would retain any coherent statistics associated with $\cohsvec[1]{\lambda\Uppr}$, since they could still be needed for future semicoherent templates; indeed in Figure~\ref{fig:weave_template_banks} it is clear that the next semicoherent template in the bank, labeled $\semivec\lambda\Upppr$, could require coherent statistics associated with $\cohsvec[1]{\lambda\Uppr}$, since the bounding boxes of $\semivec{\lambda\Uppr}(\cohsvec[1]{\lambda\Uppr})$ and $\semivec\lambda\Upppr$ intersect.

The cache management algorithm described above is implemented in the main search loop in steps~7--11 (Figure~\ref{fig:weave_schematic}).
In step~7 the cache is interrogated for a required $\calF$-statistic value $2\calF(\cohsvec\lambda)$: if it is in the cache, it is retrieved and utilized (step~8), otherwise it is computed and inserted into the cache (step~9).
In the latter case, the cache is also checked to see if any cache items can be discarded.
Starting with step~10, cache items indexed by $\cohsvec\lambda$ are retrieved in order of ascending $\frakR(\cohsvec\lambda)$.
If $\frakR(\cohsvec\lambda) < \frakR(\semivec\lambda)$, the cache items are discarded (step~11).
Only one cache item is removed at any one time, and therefore the memory usage of the cache will either remain constant, or increase by one item per main search loop iteration.
The cache is implemented using two data structures~\cite[e.g.][]{Morin2013}: a binary heap to rank cache items by $\frakR(\cohsvec\lambda)$, and a hash table to find cache items indexed by $\cohsvec\lambda$.

\section{Models of Weave Behavior}\label{sec:models-weave-behav}

\begin{table}
\centering
\begin{tabular*}{\linewidth}{l@{\extracolsep{\fill}}r@{\extracolsep{\fill}}r@{\extracolsep{\fill}}r@{\extracolsep{\fill}}r@{\extracolsep{\fill}}r@{\extracolsep{\fill}}r@{\extracolsep{2\tabcolsep}}}
\hline\hline
\multicolumn{1}{c}{$N$} & \multicolumn{1}{c}{$\coh T$} & \multicolumn{1}{c}{$\semi T$} & \multicolumn{1}{c}{$|\{(\coh\mu\umax, \semi\mu\umax)\}|$} & \multicolumn{1}{c}{$\coh\mu\umax$} & \multicolumn{1}{c}{$\semi\mu\umax$} & \multicolumn{1}{c}{$n\uinj$} \\
\hline
228 & 25.0 & 256.0 & 56 & 0.1--1.2 & 1.5--12.0 & 111.4 \\
195 & 30.0 & 256.2 & 56 & 0.1--1.6 & 2.0--12.0 & 111.4 \\
147 & 40.0 & 256.6 & 71 & 0.1--1.5 & 4.0--24.0 & 111.4 \\
\hline\hline
\end{tabular*}
\caption{\label{tab:injection_data_details}
Details of search setups used to test model in Section~\ref{sec:mism-distr}.
Columns are (left to right): number of segments, timespan of each segment in hours, total timespan of all segments in days, number of $(\coh\mu\umax, \semi\mu\umax)$ pairs used, ranges of maximum coherent and semicoherent mismatches, average number of injections per $(\coh\mu\umax, \semi\mu\umax)$ pair.
}
\end{table}

\begin{table}
\centering
\begin{tabular*}{\linewidth}{l@{\extracolsep{\fill}}r@{\extracolsep{\fill}}r@{\extracolsep{\fill}}r@{\extracolsep{\fill}}r@{\extracolsep{\fill}}r@{\extracolsep{\fill}}r@{\extracolsep{\fill}}r@{\extracolsep{\fill}}r@{\extracolsep{0pt}}l@{\extracolsep{\fill}}r@{\extracolsep{2\tabcolsep}}}
\hline\hline
\multicolumn{1}{c}{$N$} & \multicolumn{1}{c}{$\coh T$} & \multicolumn{1}{c}{$\semi T$} & \multicolumn{1}{c}{$\coh\mu\umax$} & \multicolumn{1}{c}{$\semi\mu\umax$} & \multicolumn{1}{c}{$K$} & \multicolumn{1}{c}{$|\{k\}|$} & \multicolumn{1}{c}{$\Delta f$} & \multicolumn{2}{c}{$\Delta \dot f$} & \multicolumn{1}{c}{$\calF\text{~alg.}$} \\
\hline
10 & 25.0 & 10.5 & 0.1 & 0.1 & 100 & 24 & 0.1 & $4$ & $.5{\times}10^{-9}$ & R \\
10 & 25.0 & 10.5 & 0.1 & 0.1 & 100 & 24 & 0.5 & $1$ & $.0{\times}10^{-9}$ & R \\
10 & 25.0 & 10.5 & 0.1 & 0.1 & 100 & 24 & 0.1 & $5$ & $.0{\times}10^{-9}$ & D \\
10 & 25.0 & 10.5 & 0.1 & 0.2 & 24 & 24 & 0.1 & $3$ & $.6{\times}10^{-9}$ & R \\
10 & 25.0 & 10.5 & 0.1 & 0.2 & 24 & 24 & 0.5 & $8$ & $.0{\times}10^{-10}$ & R \\
10 & 25.0 & 10.5 & 0.1 & 0.2 & 24 & 24 & 0.1 & $4$ & $.0{\times}10^{-9}$ & D \\
29 & 25.0 & 33.8 & 0.1 & 0.5 & 100 & 24 & 0.1 & $8$ & $.1{\times}10^{-10}$ & R \\
29 & 25.0 & 33.8 & 0.1 & 0.5 & 100 & 24 & 0.5 & $1$ & $.8{\times}10^{-10}$ & R \\
29 & 25.0 & 33.8 & 0.1 & 0.5 & 100 & 24 & 0.1 & $9$ & $.0{\times}10^{-10}$ & D \\
29 & 25.0 & 33.8 & 0.1 & 0.5 & 500 & 24 & 0.1 & $9$ & $.0{\times}10^{-10}$ & R \\
29 & 25.0 & 33.8 & 0.1 & 0.5 & 500 & 24 & 0.5 & $2$ & $.0{\times}10^{-10}$ & R \\
29 & 25.0 & 33.8 & 0.1 & 0.5 & 500 & 24 & 0.1 & $1$ & $.0{\times}10^{-9}$ & D \\
90 & 25.0 & 105.2 & 0.01 & 0.8 & 2700 & 24 & 0.1 & $1$ & $.8{\times}10^{-10}$ & R \\
90 & 25.0 & 105.2 & 0.01 & 0.8 & 2700 & 24 & 0.5 & $4$ & $.0{\times}10^{-11}$ & R \\
90 & 25.0 & 105.2 & 0.01 & 0.8 & 2700 & 24 & 0.1 & $2$ & $.0{\times}10^{-10}$ & D \\
90 & 25.0 & 105.2 & 0.1 & 0.8 & 1000 & 24 & 0.1 & $2$ & $.7{\times}10^{-9}$ & R \\
90 & 25.0 & 105.2 & 0.1 & 0.8 & 1000 & 24 & 0.5 & $6$ & $.0{\times}10^{-10}$ & R \\
90 & 25.0 & 105.2 & 0.1 & 0.8 & 1000 & 24 & 0.1 & $3$ & $.0{\times}10^{-9}$ & D \\
228 & 25.0 & 256.0 & 0.1 & 2 & 5000 & 24 & 0.1 & $8$ & $.1{\times}10^{-10}$ & R \\
228 & 25.0 & 256.0 & 0.1 & 2 & 5000 & 24 & 0.5 & $1$ & $.8{\times}10^{-10}$ & R \\
228 & 25.0 & 256.0 & 0.1 & 2 & 5000 & 24 & 0.1 & $9$ & $.0{\times}10^{-10}$ & D \\
228 & 25.0 & 256.0 & 0.1 & 2 & 5000 & 24 & 0.1 & $5$ & $.4{\times}10^{-10}$ & R \\
228 & 25.0 & 256.0 & 0.1 & 2 & 5000 & 24 & 0.5 & $1$ & $.2{\times}10^{-10}$ & R \\
228 & 25.0 & 256.0 & 0.1 & 2 & 5000 & 24 & 0.1 & $6$ & $.0{\times}10^{-10}$ & D \\
\hline\hline
\end{tabular*}
\caption{\label{tab:weave_test_runs_details}
Details of search setups used to test models in Sections~\ref{sec:number-templates}--~\ref{sec:input-data-bandwidth}.
Columns are (left to right): number of segments, timespan of each segment in hours, total timespan of all segments in days, maximum coherent and semicoherent mismatches, number of patches used to partition sky, number of sky patches used to test models, range of frequency parameter space in Hz, range of spindown parameter space in Hz/s, $\calF$-statistic algorithm (D=demodulation, R=resampling).
}
\end{table}

This section presents semi-analytic models of the Weave implementation.
It greatly facilitates the practical usage of any search method if its behavior can be characterized \emph{a priori} as much as possible using a computationally-cheap model.
For example, a model of the distribution of $\calF$-statistic mismatches (Section~\ref{sec:mism-distr}) permits the estimation of the sensitivity of a particular search setup~\cite{Wett2012:EsSnWdpSrGrvP} which in turn allows the setup to be optimized so as to maximize sensitivity~\cite{PrixShal2012:SCntGrvWOpStMFCmC}.
Similarly, models of the number of coherent and semicoherent templates (Section~\ref{sec:number-templates}) and computational cost (Section~\ref{sec:computational-cost}) allow the parameters of the optimal search setup to be estimated~\cite{PrixShal2012:SCntGrvWOpStMFCmC}.
The memory usage (Section~\ref{sec:memory-usage}) and input data bandwidth (Section~\ref{sec:input-data-bandwidth}) required by the implementation are also important properties when implementing a search pipeline.

Each model presented in this section is implemented as an Octave~\cite{Octave2015} script, and is freely available as part of the OctApps~\cite{WettEtAl2018:OcLOFnCntGrvDAn} script library.

\subsection{Distribution of $\calF$-statistic mismatches}\label{sec:mism-distr}

\begin{figure*}
\centering
\subfloat[]{\includegraphics[width=0.33\linewidth]{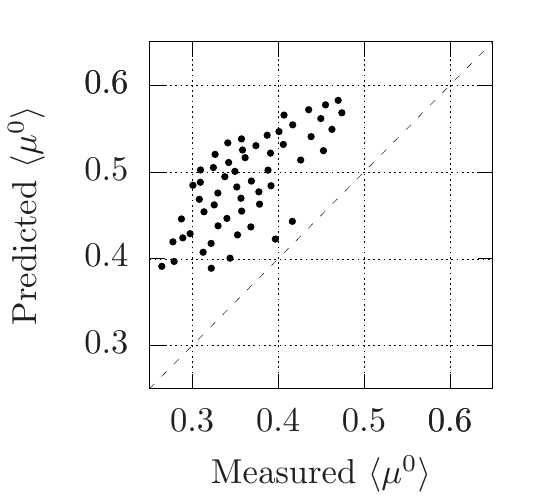}\label{fig:mismatch_model_mean_1}}
\subfloat[]{\includegraphics[width=0.33\linewidth]{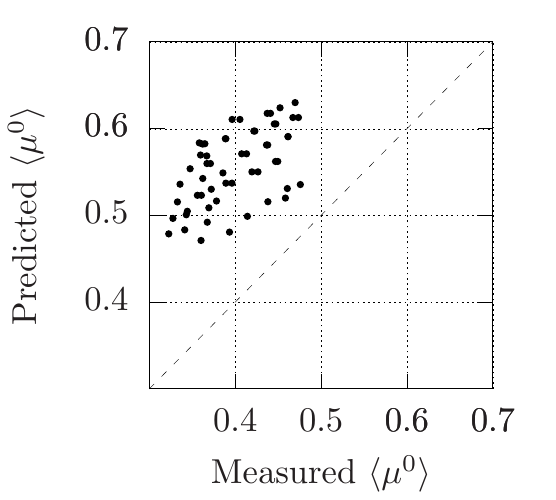}\label{fig:mismatch_model_mean_2}}
\subfloat[]{\includegraphics[width=0.33\linewidth]{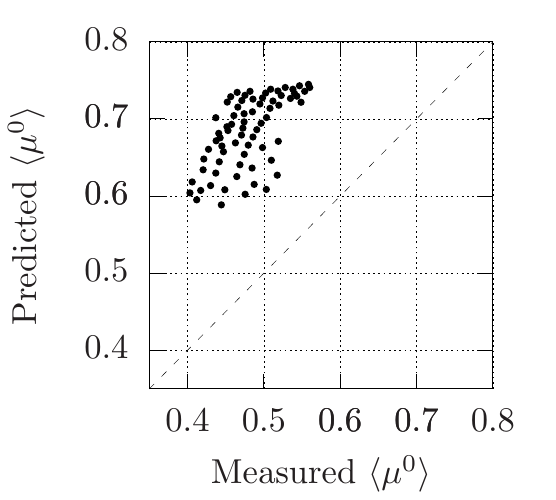}\label{fig:mismatch_model_mean_3}}\\
\subfloat[]{\includegraphics[width=0.33\linewidth]{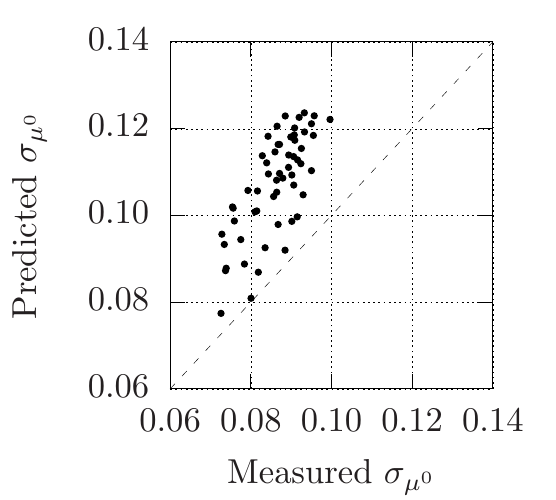}\label{fig:mismatch_model_stdv_1}}
\subfloat[]{\includegraphics[width=0.33\linewidth]{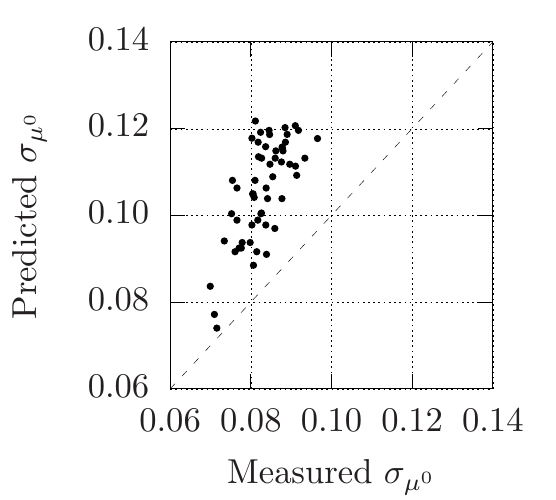}\label{fig:mismatch_model_stdv_2}}
\subfloat[]{\includegraphics[width=0.33\linewidth]{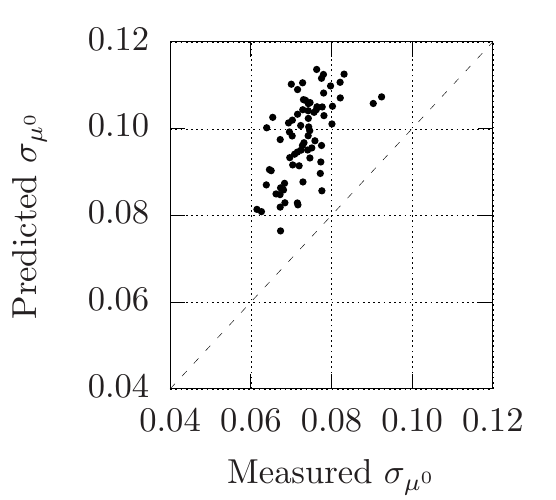}\label{fig:mismatch_model_stdv_3}}
\caption{\label{fig:mismatch_model}
Predicted means (top row) and standard deviations (bottom row) of the distributions of $\calF$-statistic mismatches, against their measured values, for (left column to right column) the 3 search setups listed in Table~\ref{tab:injection_data_details}.
The dotted line denotes equality between predicted and measured values.
}
\end{figure*}

The distribution of the mismatch between the $\calF$-statistic computed at an exact signal location, and at the nearest point in the Weave semicoherent template bank, gives an idea of the expected loss in signal-to-noise ratio due to the necessary coarseness of the template bank.
Figure~\ref{fig:mismatch_model} plots the predicted means and standard deviations of Weave $\calF$-statistic mismatch distributions, against their measured values, for a variety of setups given in Table~\ref{tab:injection_data_details}.
The distributions were measured using software injection studies, where relatively strong ($h_0 / \sqrt{S_h} \gtrsim 70 \text{Hz}^{1/2}$) simulated signals are added to Gaussian-distributed noise and then searched for using \lalapps{Weave}.

The predicted means and standard deviations are from the model presented in~\cite{Wett2016:EmExRVlPrmMASGrvP}, and are generally conservative: Figure~\ref{fig:mismatch_model} shows that the model generally overestimates the mean $\calF$-statistic mismatch by $\sim 0.13$ (Figure~\ref{fig:mismatch_model_mean_1}) to $\sim 0.20$ (Figure~\ref{fig:mismatch_model_mean_3}); and the predicted standard deviations imply slightly broader distributions than were measured.
As explored in~\cite{Wett2016:EmExRVlPrmMASGrvP}, the relationship between the maximum mismatches of the coherent and semicoherent template banks (which are inputs to \lalapps{Weave}) and the $\calF$-statistic mismatch distribution (which is output by \lalapps{Weave}) is difficult to model when the former are large e.g.\ $\gtrsim 1$.

In addition, an optimization implemented in Weave but not accounted for in the model of~\cite{Wett2016:EmExRVlPrmMASGrvP} complicates the picture: the coherent and semicoherent template banks are constructed to have equally-spaced templates in the frequency parameter $f$.
This permits (in step~9 of Figure~\ref{fig:weave_schematic}) the simultaneous computation of a series of $2\calF$ values at equally-spaced values of $f$ across the frequency parameter space, which can be performed efficiently using Fast Fourier Transform-based algorithms (see Section~\ref{sec:computational-cost}).
The construction of equal-frequency-spacing coherent and semicoherent template banks is performed by first constructing each bank independently, and then reducing the frequency spacing in all banks to that of the smallest frequency spacing in any bank.
This construction will always reduce the maximum possible mismatch in each grid, but never increase it, and so we would expect the mean $\calF$-statistic mismatch measured by Weave to be smaller than that predicted by the model of~\cite{Wett2016:EmExRVlPrmMASGrvP}.

The model of~\cite{Wett2016:EmExRVlPrmMASGrvP} is implemented in the OctApps script \octave{WeaveFstatMismatch}.

\subsection{Number of templates}\label{sec:number-templates}

\begin{figure}
\centering
\subfloat[]{\includegraphics[width=\linewidth]{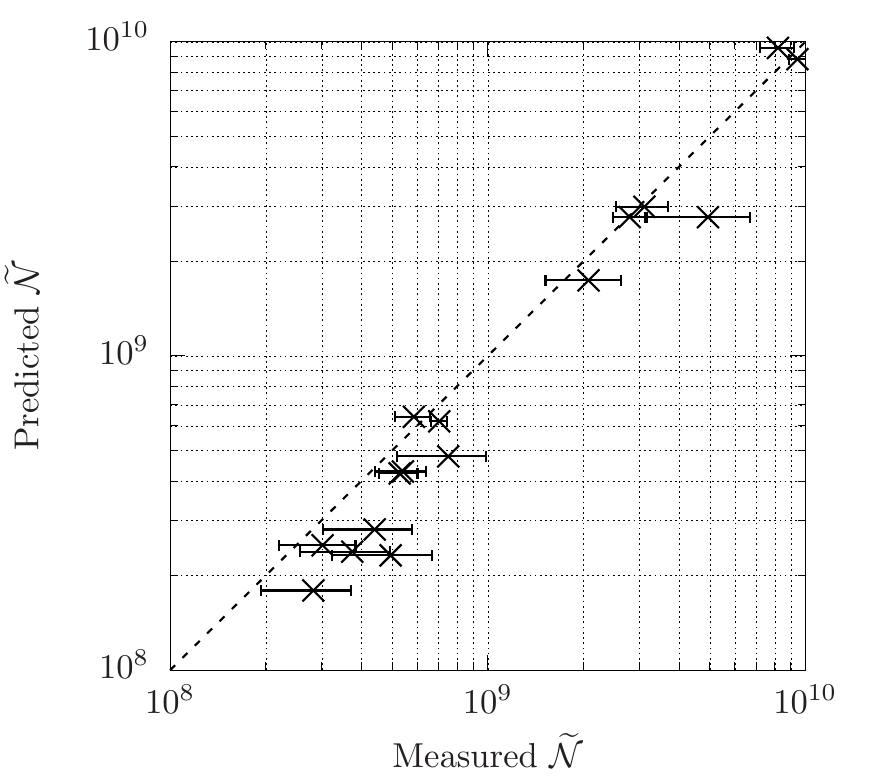}\label{fig:template_model_coh_Nt}}\\
\subfloat[]{\includegraphics[width=\linewidth]{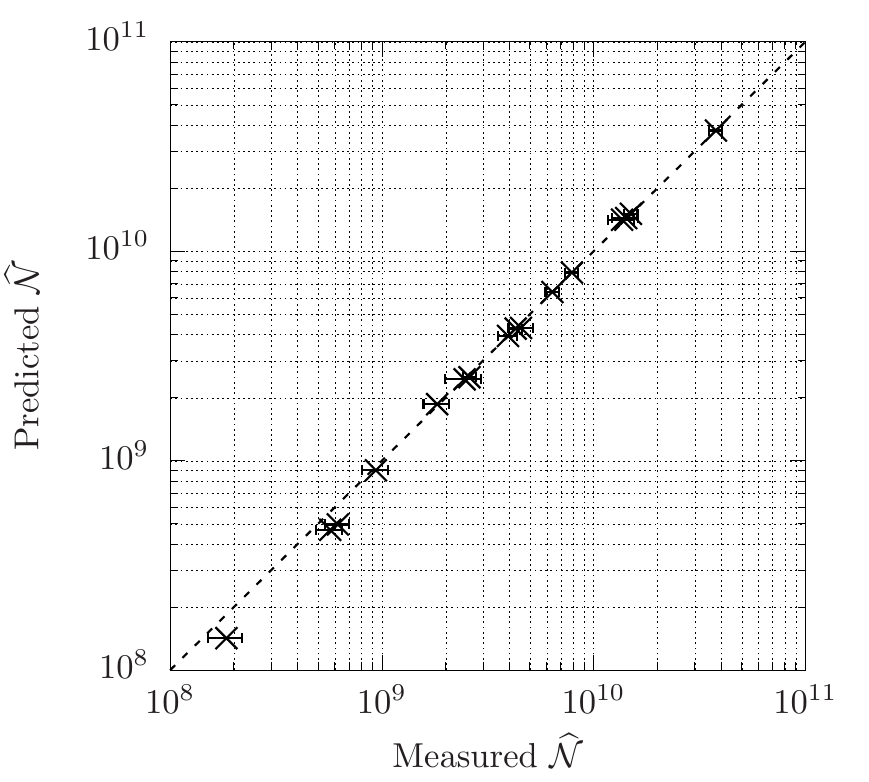}\label{fig:template_model_semi_Nt}}
\caption{\label{fig:template_model}
Predicted number of \protect\subref{fig:template_model_coh_Nt}~coherent templates [Eq.~\eqref{eq:num-coh-templates}] and \protect\subref{fig:template_model_semi_Nt}~semicoherent templates [Eq.~\eqref{eq:num-semi-templates}], against their measured values.
The error bars denote the standard deviations of measured $\coh\calN$ and $\semi\calN$ when averaged over different sky patches.
The dotted line denotes equality between predicted and measured values.
The search setups used in this Figure are listed in Table~\ref{tab:weave_test_runs_details}.
}
\end{figure}

Since the Weave coherent and semicoherent template banks are constructed using lattices (see Section~\ref{sec:optim-templ-plac}), the number of templates in each is estimated starting from the formula~\cite[e.g.][]{Prix2007:TmpSrGrvWEfLCFPrS,Wett2014:LTmPlcChASrGrvP}
\begin{equation}
\label{eq:num-templates}
\calN = \theta \mu\umax^{-n/2} \sqrt{\det \mat g} \; \calV \,,
\end{equation}
where $\cal V$ is the volume of the $n$-dimensional parameter space, $\mat g$ the parameter-space metric, and $\mu\umax$ the maximum mismatch.
The \emph{normalized thickness} $\theta$ is a property of the particular lattice used to generate the template bank~\cite[e.g.]{ConwaySloane1988}.

The parameter-space volume is given explicitly by the following expressions:
\begin{align}
\label{eq:param-vol}
\calV &= \calV_{n\ua n\ub} \prod_{s=0}^{s\umax} \calV_{\ndot f} \,, \\
\begin{split}
\label{eq:param-vol-sky}
\calV_{n\ua n\ub} &= \max \{ \beta\ua \beta\ub, \calV_{n\ua n\ub}\Upr \} \,, \\
\calV_{n\ua n\ub}\Upr &= 2 ( \pi + 4\beta\ub + \beta\ua \beta\ub ) \\
&\quad\times \begin{cases}
\frac{ (\alpha\umax - \alpha\umin) (\sin\delta\umax - \sin\delta\umin) }{ 4\pi } & \text{or} \\
\frac{1}{ K } \,,
\end{cases}
\end{split} \\
\label{eq:param-vol-fspin}
\calV_{\ndot f} &= \beta_{\ndot f} + \ndot f\umax - \ndot f\umin \,.
\end{align}
Here, $\vec\beta$ is the vector whose components are the extents of the bounding box of $\mat g$ in each dimension; it is used to ensure that the volume of the parameter space in each dimension is not smaller than the extent of a single template.
In Eq.~\eqref{eq:param-vol-sky}, the volume of the sky parameter space may be specified either by a rectangular patch $[\alpha\umin, \alpha\umax] \otimes [\delta\umin, \delta\umax]$, or by the number $K$ of equal-size sky patches (see Section~\ref{sec:overview}).

Finally, the total number of coherent and semicoherent templates, $\coh\calN$ and $\semi\calN$ respectively, are given by:
\begin{align}
\label{eq:num-coh-templates}
\coh\calN &= 1.436 \sum_{\ell=0}^{N-1} \theta \coh\mu\umax^{-n/2} \sqrt{\det \cohsmat g} \; \coh\calV \,, \\
\label{eq:num-semi-templates}
\semi\calN &= \theta \semi\mu\umax^{-n/2} \sqrt{\det \semimat g} \; \semi\calV \,.
\end{align}
The numerical prefactor on the right-hand side of Eq.~\eqref{eq:num-coh-templates} is chosen to better match $\coh\calN$ to the number of coherent templates actually computed by \lalapps{Weave}: the coherent parameter space is augmented with additional padding along its boundaries to ensure that it encloses the semicoherent parameter space, i.e.\ that it includes a nearest neighbor for every $\semivec\lambda$.

Equations~\eqref{eq:num-coh-templates} and~\eqref{eq:num-semi-templates} are used to predict the number of templates computed by \lalapps{Weave} for a variety of search setups detailed in Table~\ref{tab:weave_test_runs_details}.
Figure~\ref{fig:template_model} plots the predicted $\coh\calN$ and $\semi\calN$ against the values measured by running \lalapps{Weave}.
Reasonable agreement is achieved between predicted and measured $\coh\calN$ (Figure~\ref{fig:template_model_coh_Nt}): while Eq.~\eqref{eq:num-coh-templates} sometimes underestimates the number of coherent templates, it rarely does so by more than a factor of a few.
Better agreement is seen between predicted and measured $\semi\calN$ (Figure~\ref{fig:template_model_semi_Nt}).

Equations~\eqref{eq:num-coh-templates} and~\eqref{eq:num-semi-templates} are implemented in the OctApps script \octave{WeaveTemplateCount}.

\subsection{Computational cost}\label{sec:computational-cost}

\begin{table}
\centering
\begin{tabular*}{\linewidth}{l@{\extracolsep{\fill}}r@{\extracolsep{\fill}}r@{\extracolsep{2\tabcolsep}}}
\hline\hline
Fundamental Timing Constant & \multicolumn{2}{c}{Representative Value / s} \\
 & Demodulation & Resampling \\
\hline
$\tau\Ueff\uFstat$ &  $2.5{\times}10^{-6}$ &  ($1.5$--$4$)${\times}10^{-7}$ \\
$\tau\uiter$ & \multicolumn{2}{c}{ $1.4{\times}10^{-10}$} \\
$\tau\uquery$ & \multicolumn{2}{c}{ $8.6{\times}10^{-11}$} \\
$\tau\usemimeantwof$ & \multicolumn{2}{c}{ $8.3{\times}10^{-10}$} \\
$\tau\usemisegsumtwof$ & \multicolumn{2}{c}{ $7.3{\times}10^{-10}$} \\
$\tau\usemilogtenbsgl$ & \multicolumn{2}{c}{ $9.9{\times}10^{-9}$} \\
$\tau\uoutput$ & \multicolumn{2}{c}{ $7.9{\times}10^{-10}$} \\
\hline\hline
\end{tabular*}
\caption{\label{tab:fundamental_timing_constants}
Representative values of the fundamental timing constants of the computational cost model detailed in Section~\ref{sec:computational-cost}.
These values were computed by running \lalapps{Weave} on a computer cluster of Intel Xeon E5-2658V4 processors running at 2.30~GHz.
Some values are specific to the search setups detailed in Table~\ref{tab:weave_test_runs_details}.
}
\end{table}

\begin{figure*}
\centering
\subfloat[]{\includegraphics[width=0.33\linewidth]{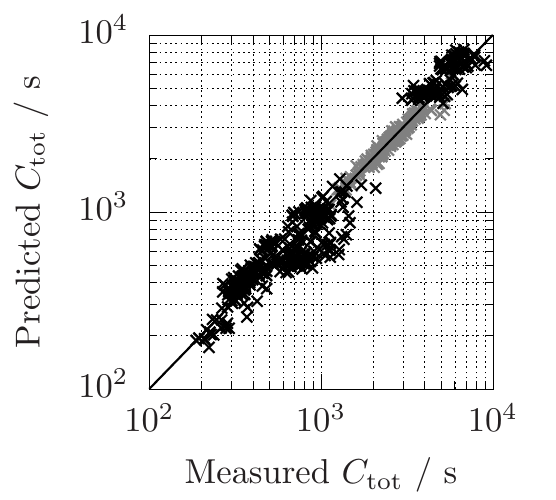}\label{fig:timing_model_total}}
\subfloat[]{\includegraphics[width=0.33\linewidth]{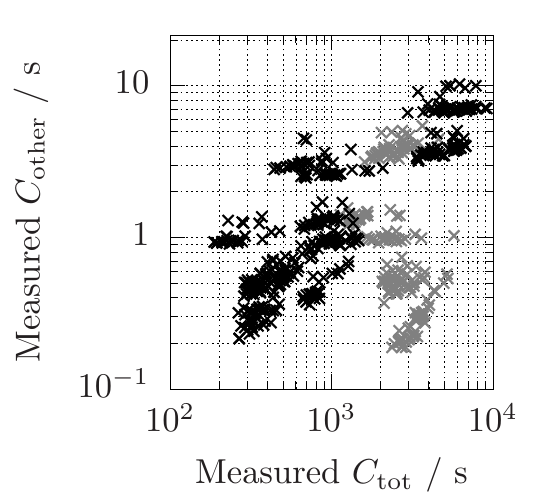}\label{fig:timing_model_total_other}}
\subfloat[]{\includegraphics[width=0.33\linewidth]{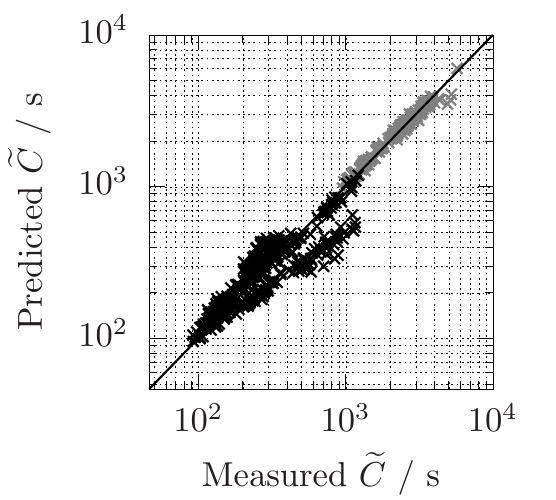}\label{fig:timing_model_coh_coh2f}}\\
\subfloat[]{\includegraphics[width=0.33\linewidth]{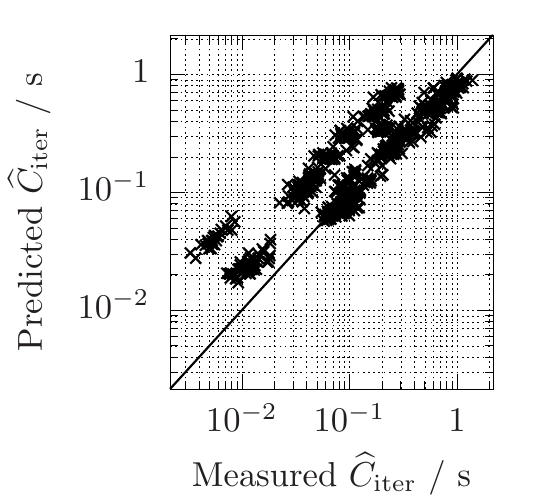}\label{fig:timing_model_iter}}
\subfloat[]{\includegraphics[width=0.33\linewidth]{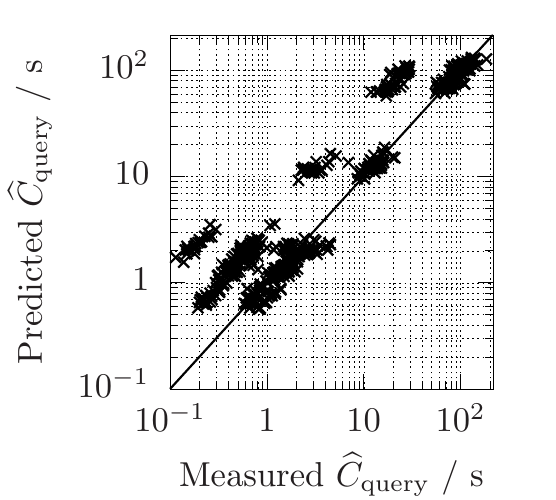}\label{fig:timing_model_query}}
\subfloat[]{\includegraphics[width=0.33\linewidth]{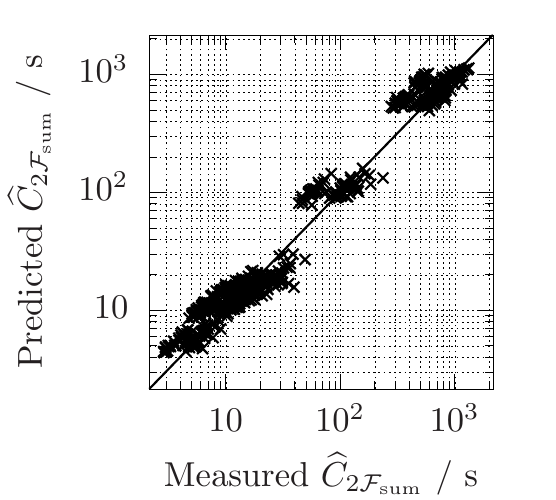}\label{fig:timing_model_semiseg_sum2f}}\\
\subfloat[]{\includegraphics[width=0.33\linewidth]{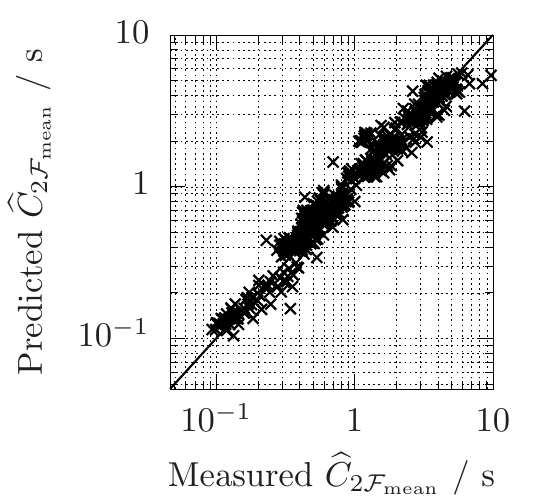}\label{fig:timing_model_semi_mean2f}}
\subfloat[]{\includegraphics[width=0.33\linewidth]{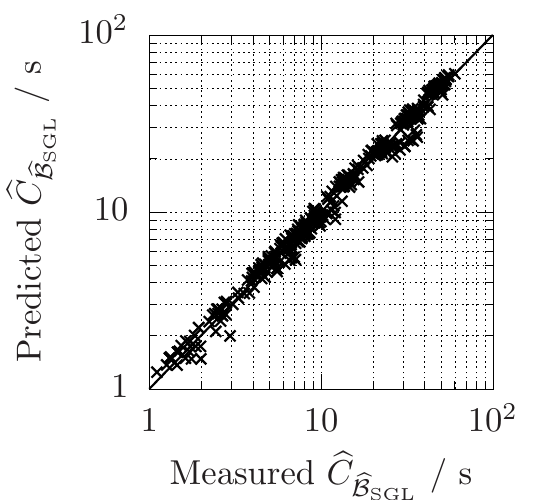}\label{fig:timing_model_semi_log10bsgl}}
\subfloat[]{\includegraphics[width=0.33\linewidth]{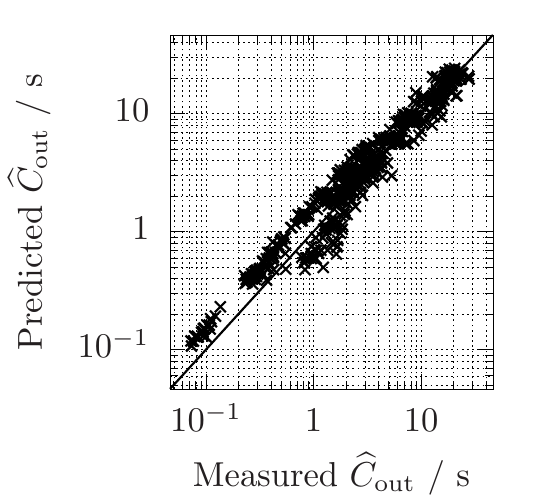}\label{fig:timing_model_output}}
\caption{\label{fig:timing_model}
\protect\subref{fig:timing_model_total} Predicted vs.\ measured total computational cost $C\utotal$ [Eq.~\eqref{eq:comp-cost-tot}].
\protect\subref{fig:timing_model_total_other} Measured unmodeled computational cost $C\uother$ [Eq.~\eqref{eq:comp-cost-tot}] vs.\ measured total computational cost $C\utotal$.
\protect\subref{fig:timing_model_coh_coh2f}--\protect\subref{fig:timing_model_output} Predicted vs.\ measured values of the coherent computational cost $\coh C$ of Eq.~\eqref{eq:comp-cost-coh}, and of the components of the semicoherent computational cost $\semi C$ of Eq.~\eqref{eq:comp-cost-semi}.
In \protect\subref{fig:timing_model_total}--\protect\subref{fig:timing_model_coh_coh2f}, grey crosses and black crosses are used to distinguish values computed using the demodulation and resampling $\calF$-statistic algorithms respectively; otherwise black crosses are used for values which are independent of the choice of $\calF$-statistic algorithm.
The search setups used in this Figure are listed in Table~\ref{tab:weave_test_runs_details}.
}
\end{figure*}

The total computational cost $C\utotal$ of a particular search setup may be modeled in terms of the number of coherent $\coh\calN$ and semicoherent $\semi\calN$ templates (see Section~\ref{sec:number-templates}), the number of segments $N$ and number of detectors $N\Udet$.
Following~\cite{PrixShal2012:SCntGrvWOpStMFCmC} we write
\begin{equation}
\label{eq:comp-cost-tot}
\begin{split}
C\utotal(\coh\calN, \semi\calN, N, N\Udet) &= \coh C(\coh\calN, N\Udet) \\ &\quad + \semi C(\semi\calN, N, N\Udet) + C\uother \,,
\end{split}
\end{equation}
where $\coh C$ and $\semi C$ denote the computational cost of the coherent and semicoherent stages of the search method respectively, and $C\uother$ denotes any unmodeled computational costs.

The computational cost model takes as input \emph{fundamental timing constants} which give the time taken to complete certain fundamental computations.
Their values are highly dependent on various properties of the computer hardware used to run \lalapps{Weave}, such as the processor speed and cache sizes, as well as what other programs were using the computer hardware at the same time as \lalapps{Weave}.
Some values are also specific to the search setups detailed in Table~\ref{tab:weave_test_runs_details}.
For the interest of the reader, Table~\ref{tab:fundamental_timing_constants} lists representative values of the fundamental timing constants obtained on a particular computer cluster.

The coherent cost $\coh C$ is simply the cost of computing the $\calF$-statistic (step~9 of Figure~\ref{fig:weave_schematic}):
\begin{equation}
\label{eq:comp-cost-coh}
\coh C(\coh\calN, N\Udet) = \coh\calN N\Udet \tau\Ueff\uFstat(\Delta f, \coh T, \calF\text{~alg.}) \,.
\end{equation}
The fundamental timing constant $\tau\Ueff\uFstat$ gives the time taken to compute the $\calF$-statistic per template and per detector, and is further described in~\cite{Prix2017:ChTMmrFsImpLA}.
Its value depends primarily upon the range of the frequency parameter space $\Delta f$, the coherent segment length $\coh T$, and the algorithm used to compute the $\calF$-statistic.
Choices for the latter are: the \emph{resampling} algorithm~\cite[e.g.][]{JaraEtAl1998:DAnGrvSgSpNSSDtc,PateEtAl2010:ImpBrRsCnWSGrvWD}, which computes the $\calF$-statistic over a wide band of frequencies efficiently using the Fast Fourier Transform, and is generally used to performing an initial wide-parameter-space search; and the \emph{demodulation} algorithm of~\cite{WillSchu2000:EfMFlAlDCntGrvWS}, which uses a Dirichlet kernel to compute the $\calF$-statistic more efficiently at a single frequency or over a narrow frequency band, and is therefore used to perform follow-up searches of localized parameter spaces around interesting candidates.
The additional cost of managing the cache of computed $\calF$-statistic values (steps~8, 10, and~11) is amortized into $\coh C$.

The semicoherent cost
\begin{equation}
\label{eq:comp-cost-semi}
\begin{split}
\semi C(\semi\calN, N, N\Udet) &= \semi C\uiter(\semi\calN) + \semi C\uquery(\semi\calN, N) \\ &\quad + \semi C\usemisegsumtwof(\semi\calN, N, N\Udet) \\ &\quad + \semi C\usemimeantwof(\semi\calN) + \semi C\usemilogtenbsgl(\semi\calN) \\ &\quad + \semi C\uoutput(\semi\calN)
\end{split}
\end{equation}
has a number of components:
\begin{enumerate}[(i)]
\item $\semi C\uiter$ is the cost of iterating over the semicoherent template bank (steps~5 and~16 of Figure~\ref{fig:weave_schematic});
\item $\semi C\uquery$ is the cost of finding the nearest templates in the coherent template banks (step~6 and~13) and of interrogating the cache of computed $\calF$-statistic values (step~7);
\item $\semi C\usemisegsumtwof$ is the cost of computing $2\calF\usum$ and, if required, $2\calF^X\usum$ (step~12);
\item $\semi C\usemimeantwof$ is the cost of computing $2\calF\umean$ (step~14);
\item $\semi C\usemilogtenbsgl$ is the cost of computing $\semi\calB\uSGL$, if required (step~14); and
\item $\semi C\uoutput$ is the cost of adding candidates to toplists (step~15).
\end{enumerate}
These components of $\semi C$ are further defined in terms of $\semi\calN$, $N$, $N\Udet$, and various fundamental timing constants (see Table~\ref{tab:fundamental_timing_constants}) as follows:
\begin{align}
\label{eq:comp-cost-semi-detail}
\semi C\uiter(\semi\calN) &= \semi\calN \tau\uiter \,; \\
\semi C\uquery(\semi\calN, N) &= \semi\calN N \tau\uquery \,; \\
\semi C\usemisegsumtwof(\semi\calN, N, N\Udet) &= \semi\calN (N - 1) \tau\usemisegsumtwof \nonumber \\
&\quad\times \begin{cases}
1 + N\Udet & \text{if~} 2\calF^X\usum \text{~required} \,, \\
1 & \text{otherwise} \,;
\end{cases} \\
\semi C\usemimeantwof(\semi\calN) &= \semi\calN \tau\usemimeantwof \,; \\
\semi C\usemilogtenbsgl(\semi\calN) &= \semi\calN \tau\usemilogtenbsgl \,; \\
\semi C\uoutput(\semi\calN) &= \semi\calN \tau\uoutput \times \text{number of toplists} \,.
\end{align}

Figure~\ref{fig:timing_model} compares the computational cost model of Eqs.~\eqref{eq:comp-cost-tot}--\eqref{eq:comp-cost-semi-detail} against the measured computational cost of \lalapps{Weave} (see Table~\ref{tab:fundamental_timing_constants}), using the search setups detailed in Table~\ref{tab:weave_test_runs_details}.
The total computational cost of \lalapps{Weave} is generally well-modeled (Figure~\ref{fig:timing_model_total}) and the unmodeled component of the measured computational cost is low (Figure~\ref{fig:timing_model_total_other}).
The coherent computational cost $\coh C$ of Eq.~\eqref{eq:comp-cost-coh} and the components of the semicoherent cost $\semi C$ of Eq.~\eqref{eq:comp-cost-semi} are also in good agreement (Figures~\ref{fig:timing_model_coh_coh2f}--\ref{fig:timing_model_output}).

Equations~\eqref{eq:comp-cost-tot}--\eqref{eq:comp-cost-semi-detail} are implemented in the OctApps script \octave{WeaveRunTime}.

\subsection{Memory usage}\label{sec:memory-usage}

\begin{figure}
\centering
\subfloat[]{\includegraphics[width=\linewidth]{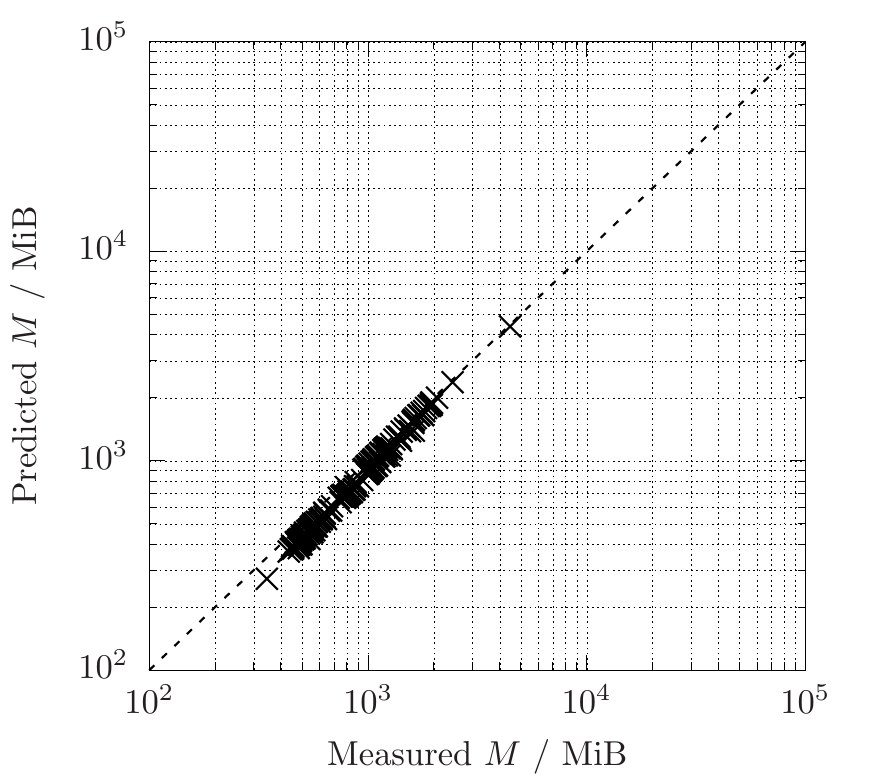}\label{fig:memory_model_demod}}\\
\subfloat[]{\includegraphics[width=\linewidth]{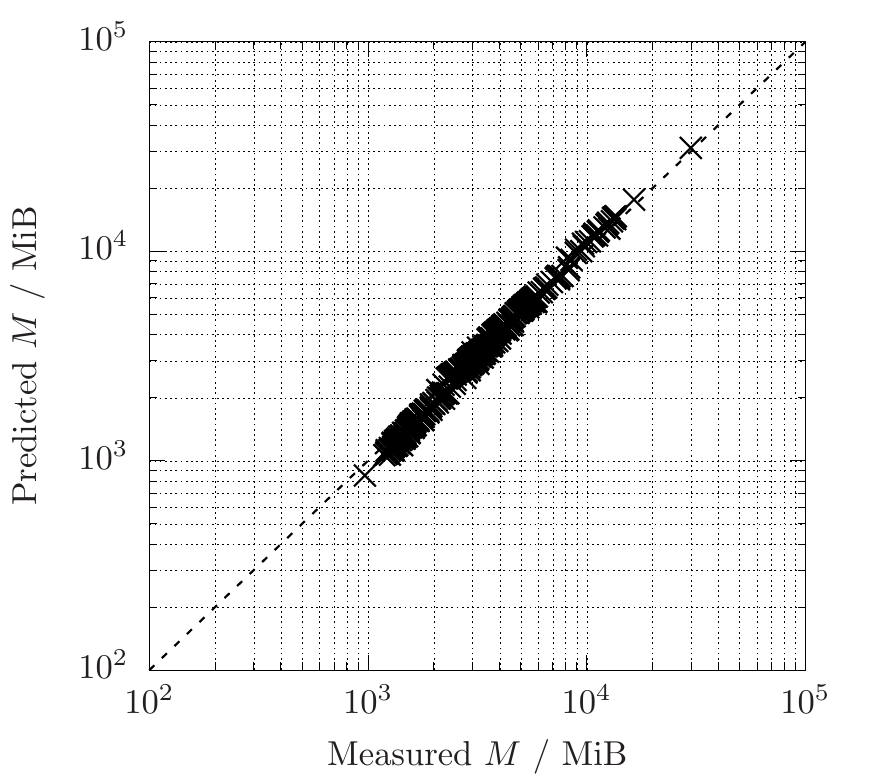}\label{fig:memory_model_resamp}}\\
\caption{\label{fig:memory_model}
Predicted memory usage of \lalapps{Weave}, against its measured memory usage, when using the \protect\subref{fig:memory_model_demod} demodulation and \protect\subref{fig:memory_model_resamp} resampling $\calF$-statistic algorithms.
The search setups used in this Figure are listed in Table~\ref{tab:weave_test_runs_details}.
}
\end{figure}

The memory usage $M$ of \lalapps{Weave} is modeled by
\begin{equation}
\label{eq:mem-usage-tot}
M = M\uFstat + M\ucache \,.
\end{equation}
The first term on the right-hand side, $M\uFstat$, is the memory usage of the $\calF$-statistic algorithm (which includes the gravitational-wave detector data) and is further described in~\cite{Prix2017:ChTMmrFsImpLA}.
The second term, $M\ucache$, is the memory usage of the cache of computed $\calF$-statistic values, and is further given by
\begin{equation}
\label{eq:mem-usage-cache}
M\ucache = N\ucache\Umax m\utwoF \begin{cases}
1 + N\Udet & \text{if~} 2\calF^X\usum \text{~reqd.} \,, \\
1 & \text{otherwise} \,,
\end{cases} \\
\end{equation}
where $N\ucache\Umax$ is the maximum size of the cache (across all segments), and $m\utwoF \equiv 4{\times}2^{-20}$~MiB (mebibytes) is the memory required to store one $2\calF$ value as a 4-byte single precision floating-point number.
The maximum cache $N\ucache\Umax$ cannot easily be predicted from first principles, i.e.\ given the search setup, parameter space, and other input arguments to \lalapps{Weave}.
Instead, it is measured by running \lalapps{Weave} in a special mode which simulates the performance of the cache but without computing any $\calF$-statistic or derived values; essentially it follows Figure~\ref{fig:weave_schematic} but with the first part of step~9, step~12, and step~14 omitted.

Figure~\ref{fig:memory_model} plots the predicted memory usage of Eqs.~\eqref{eq:mem-usage-tot} and~\eqref{eq:mem-usage-cache} against the measured memory usage of \lalapps{Weave}, using the search setups detailed in Table~\ref{tab:weave_test_runs_details}.
The $\calF$-statistic is computed using both the resampling and demodulation algorithms: in the former case, both $\calF\umean$ and $\semi\calB\uSGL$ are computed, thereby triggering the first case in Eq.~\eqref{eq:mem-usage-cache}; in the latter case, only $\calF\umean$ is computed, thereby triggering the second case in Eq.~\eqref{eq:mem-usage-cache}.
Good agreement between predicted and measured memory usage is seen for both algorithms.

Equations~\eqref{eq:mem-usage-tot} and~\eqref{eq:mem-usage-cache} are also implemented in the OctApps script \octave{WeaveRunTime}.

\subsection{Input data bandwidth}\label{sec:input-data-bandwidth}

\begin{figure}
\centering
\includegraphics[width=\linewidth]{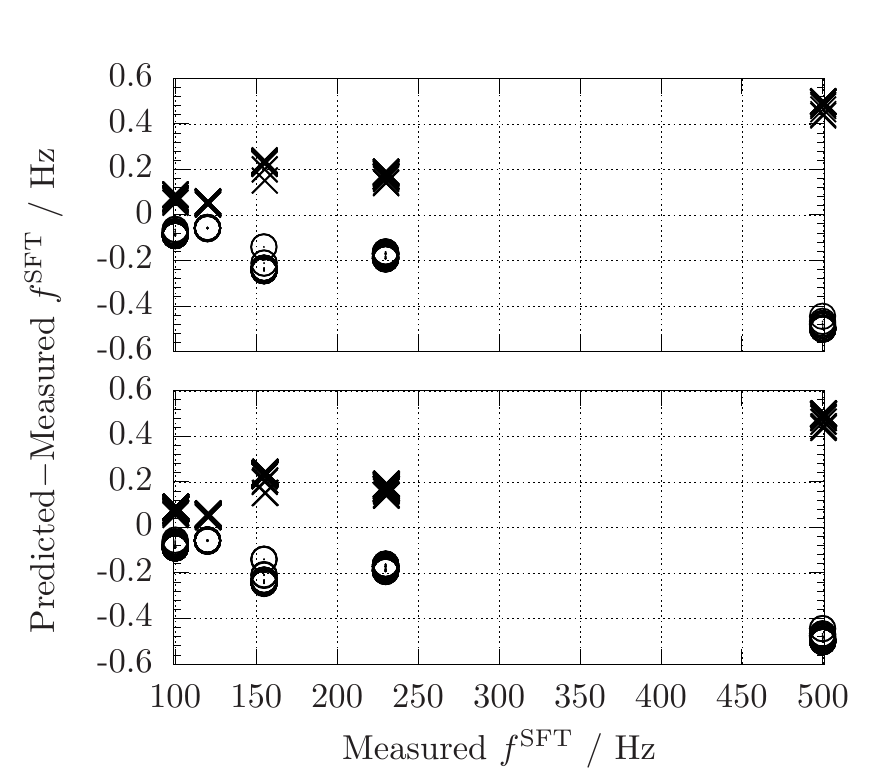}
\caption{\label{fig:input_sft_band_model}
Difference between predicted and measured $f\USFT\umin$ (circles) and $f\USFT\umax$ (crosses), against the measured $f\USFT\umin$ and $f\USFT\umax$ respectively, for the demodulation (top plot) and resampling (bottom plot) $\calF$-statistic algorithms.
The search setups used in this Figure are listed in Table~\ref{tab:weave_test_runs_details}.
}
\end{figure}

Our final Weave model concerns what bandwidth of the input gravitational-wave detector data is required to search a given frequency range.
For most continuous-wave search pipelines, short (typically 1800~s) contiguous segments of gravitational-wave strain data are Fourier transformed, and the resulting complex spectra stored as Short Fourier Transform (SFT) files.
A continuous-wave search of a large frequency parameter space will generally be divided into smaller jobs, with each job searching a smaller partition of the whole frequency parameter space.
Each job therefore requires that only a small bandwidth out of the full SFT spectra be read into memory.

Given an input frequency parameter space $[f\umin, f\umax]$ and spindown parameter space $[\dot f\umin, \dot f\umax]$, predicting the bandwidth of the SFT spectra required by \lalapps{Weave} proceeds in several steps.
First, the input parameter spaces are augmented to account for extra padding of the Weave template banks:
\begin{subequations}
\label{eqs:input-data-bandwidth-1}
\begin{align}
f\Upr\umin &= (1 - x\upad) f\umin \,, & f\Upr\umax &= (1 + x\upad) f\umax \,, \\
\dot f\Upr\umin &= (1 - \dot x\upad) \dot f\umin \,, & \dot f\Upr\umax &= (1 + \dot x\upad) \dot f\umax \,,
\end{align}
\end{subequations}
where $x\upad \equiv 10^{-3}$ and $\dot x\upad \equiv 10^{-10}$ are empirically chosen.
Next, the maximum frequency range $[f\Uppr\umin, f\Uppr\umax]$ is found by evolving the frequency--spindown parameter space $[f\Upr\umin, f\Upr\umax] \otimes [ \dot f\Upr\umin, \dot f\Upr\umax]$ from the reference time $t_0$ to the start and end times of each segment, $t\Ustart\useg$ and $t\Ustop\useg$ respectively:
\begin{subequations}
\label{eqs:input-data-bandwidth-2}
\begin{align}
f\Uppr\umin &= \min \big\{ f\Upr\umin + \dot f\Upr\umin (t - t_0) \,\big|\, t \in \{t\Ustart\useg, t\Ustop\useg\}_{\ell=0}^{N-1} \big\} \,, \\
f\Uppr\umax &= \max \big\{ f\Upr\umax + \dot f\Upr\umax (t - t_0) \,\big|\, t \in \{t\Ustart\useg, t\Ustop\useg\}_{\ell=0}^{N-1} \big\}
\end{align}
\end{subequations}
Finally, the SFT bandwidth $[f\USFT\umin, f\USFT\umax]$ of the SFT spectra which is required by \lalapps{Weave} is given by:
\begin{subequations}
\label{eqs:input-data-bandwidth-3}
\begin{align}
f\USFT\umin &= (1 - x\usky) f\Uppr\umin - f\uFstat \,, \\
f\USFT\umax &= (1 + x\usky) f\Uppr\umax + f\uFstat \,.
\end{align}
\end{subequations}
The $x\usky$ enlarges $[f\Uppr\umin, f\Uppr\umax]$ to account for the maximum frequency-dependent Doppler modulation of a continuous-wave signal due to the sidereal and orbital motions of the Earth, and is given by
\begin{equation}
\label{eqs:input-data-bandwidth-4}
x\usky = \frac{2\pi}{c} \left( \frac{D\uES}{1 \text{ year}} + \frac{R\uE}{1 \text{ day}} \right) \,,
\end{equation}
where $c$ is the speed of light, $D\uES$ is the Earth--Sun distance and $R\uE$ the radius of the Earth.
Additional padding of $[f\Uppr\umin, f\Uppr\umax]$ is also required for use by the chosen $\calF$-statistic algorithm, and is given by $f\uFstat$~\cite[see][]{Prix2017:ChTMmrFsImpLA}.

Figure~\ref{fig:input_sft_band_model} compares the model of Eqs.~\eqref{eqs:input-data-bandwidth-1}--\eqref{eqs:input-data-bandwidth-4} against the behavior of \lalapps{Weave} when run with the search setups detailed in Table~\ref{tab:weave_test_runs_details}.
Note that the model satisfies
\begin{equation*}
\text{predicted } f\USFT\umin - \text{measured } f\USFT\umin \le 0 \,,
\end{equation*}
i.e.\ all circles plotted in Figure~\ref{fig:input_sft_band_model} are below the horizontal axis, and
\begin{equation*}
\text{predicted } f\USFT\umax - \text{measured } f\USFT\umax \ge 0 \,,
\end{equation*}
i.e.\ all crosses plotted in Figure~\ref{fig:input_sft_band_model} are above the horizontal axis.
The model is therefore conservative, i.e.\ it may predict a slightly larger SFT bandwidth than required, but should never predict a smaller SFT bandwidth, which would cause a fatal error in \lalapps{Weave}.
The model is generally more conservative at higher frequencies, where the Doppler modulation due to the Earth's motion is higher.

Equations~\eqref{eqs:input-data-bandwidth-1}--\eqref{eqs:input-data-bandwidth-4} are implemented in the OctApps script \octave{WeaveInputSFTBand}.

\section{Discussion}\label{sec:discussion}

This paper details the Weave implementation of a semicoherent search method for continuous gravitational waves.
It focuses on all-sky surveys for isolated continuous-wave sources, for which the parameter space is the sky position and frequency evolution of the source.
We note, however, that the implementation is in fact indifferent to the parameter space being searched, as long as the relevant constant parameter-space metric is available.
The implementation could therefore be adapted to search other parameter spaces for continuous-wave sources such as known low-mass X-ray binaries, for which the parameter space includes the evolution parameters of the binary orbit, using the metric of~\cite{LeacPrix2015:DrSCnGrvWBSPrmMOSXSn}.

There is scope to improve the semi-analytic models of the behavior of \lalapps{Weave} presented in Section~\ref{sec:models-weave-behav}.
In particular, a more accurate model of the distribution of $\calF$-statistic mismatches than that presented in Section~\ref{sec:mism-distr} would allow the sensitivity of a search to be more accurately estimated without resorting to software injection studies.
The memory model of Section~\ref{sec:memory-usage} would also be improved if the maximum cache size $N\ucache\Umax$ could be predicted from first principles.

In a forthcoming paper~\cite{WeaveVsGCTPaper} we plan to more fully characterize the performance of the Weave implementation, and compare it to an implementation of the method of~\cite{PletAlle2009:ExLrCrrDCnGrvW,Plet2010:PrmMSmSrCnGrvW} using a mock data challenge.

\acknowledgments

We thank Bruce Allen and Heinz-Bernd Eggenstein for valuable discussions.
KW is supported by ARC CE170100004.
Numerical simulations were performed on the Atlas computer cluster of the Max Planck Institute for Gravitational Physics.
This paper has document number LIGO-P1800074-v4.

\appendix

\section{Properties of equal-area sky patches}\label{sec:prop-equal-area}

The search program \lalapps{Weave} allows the sky search parameter space to be partitioned into $K$ patches, and a patch selected by an index $k$.
Tests of this feature found that, provided $K \ll \calN^1_1$ (the number of templates with just one patch), the variation in the number of templates between patches $\Delta\calN_K$ is generally small and well-approximated by
\begin{equation}
\Delta\calN_K = \begin{cases}
1.48{\times}10^{-2} + 5.35{\times}10^{-4} K & K \le 100 \,, \\
8.48{\times}10^{-2} + 4.03{\times}10^{-5} K & K > 100 \,.
\end{cases}
\end{equation}
The ratio $\sum_{k=0}^{K-1} \calN^k_K / \calN^1_1$ of the number of templates in all $K$ patches to the number of templates with just one patch is generally $\lesssim 7$\%.
The union of all templates in a set of $K$ patches also faithfully reproduces the unpartitioned template bank, i.e.\ with just one patch.

\bibliography{paper}

\end{document}